\documentclass{article}
\usepackage{arxiv}

\usepackage[utf8]{inputenc} 
\usepackage[T1]{fontenc}    
\usepackage{algpseudocode} 
\usepackage{hyperref}       
\usepackage{url}            
\usepackage{booktabs}       
\usepackage{amsfonts}       
\usepackage{nicefrac}       
\usepackage{microtype}      
\usepackage{lipsum}
\usepackage{algorithm}
\usepackage{caption}
\usepackage{graphicx}
\usepackage{amsmath}
\usepackage{tikz}

\title{Energy-Time Ptychography for one dimensional Phase Retrieval}

\author{%
  Ankita Negi\textsuperscript{1,2,*}, 
  Leon Merten Lohse\textsuperscript{1,2}, 
  Sven Velten\textsuperscript{1}, 
  Ilya Sergeev\textsuperscript{1}, 
  Olaf Leupold\textsuperscript{1}, 
  Sakshath Sadashivaiah\textsuperscript{3,4}, 
  Dimitrios Bessas\textsuperscript{5}, 
  Aleksandr Chumakhov\textsuperscript{5}, 
  Christina Brandt\textsuperscript{6}, 
  Lars Bocklage\textsuperscript{1,2}, 
  Guido Meier\textsuperscript{7,2}, 
  and Ralf R\"{o}hlsberger\textsuperscript{1,2,3,4,8}%
}

\affil{1}{Department of Photon Science, Deutsches Elektronen-Synchrotron DESY, Notkestr. 85, 22607, Hamburg, Germany}
\affil{2}{The Hamburg Centre for Ultrafast Imaging, Luruper Chaussee 149, 22761 Hamburg, Germany}
\affil{3}{Helmholtz Institute Jena, Fraunhoferstr. 8, 07743, Jena, Germany}
\affil{4}{GSI Helmholtz Centre for Heavy Ion Research, Planckstr. 1, 64291, Darmstadt, Germany}
\affil{5}{European Synchrotron Radiation Facility, F-38043 Grenoble, France}
\affil{6}{Institute for Mathematics and Informatics, University of Greifswald, Walther-Rathenau-str. 47, 17489 Greifswald, Germany}
\affil{7}{Max Planck Institute for the Structure and Dynamics of Matter, Luruper Chaussee 149, 22761 Hamburg, Germany}
\affil{8}{Institute of Optics and Quantum Electronics, University of Jena, Max-Wien-Platz 1, 07743, Jena, Germany}

\email{ankita.negi@desy.de}

\begin{document}
\maketitle

\begin{abstract}
Phase retrieval is at the heart of adaptive optics and modern high-resolution imaging. Without phase information, optical systems are limited to intensity-only measurements, hindering full reconstruction of object structures and wavefront dynamics essential for advanced applications. Here, we address a one-dimensional phase problem linking energy and time, which arises in X-ray scattering from ultrasharp nuclear resonances. We leverage the M\"{o}ssbauer effect, where nuclei scatter radiation without energy loss to the lattice, and are sensitive to their magneto-chemical environments. Rather than using traditional spectroscopy with radioactive gamma-ray sources, we measure nuclear forward scattering of synchrotron X-ray pulses in the time domain, providing superior sensitivity and faster data acquisition. Extracting spectral information from a single measurement is challenging due to the missing phase information, typically requiring extensive modeling. Instead, we use multiple energetically overlapping measurements to retrieve both the transmission spectrum and the phase of the scattering response, similar to ptychographic phase retrieval in imaging. Our robust approach can overcome the bandwidth limitations of gamma-ray sources, opening new research directions with modern X-ray sources and M\"{o}ssbauer isotopes.
\end{abstract}


\newcommand{\Fe}{{}^{57}\mathrm{Fe}}
\newcommand{\Sn}{{}^{119}\mathrm{Sn}}
\newcommand{\Eu}{{}^{151}\mathrm{Eu}}
\newcommand{\eq}[1]{Eq.~(\ref{#1})}
\newcommand{\fig}[1]{Fig.~\ref{#1}}
\newcommand{\sect}[1]{Sec.~\ref{#1}}
\newcommand{\micron}{{\mu}\mathrm{m}}
\newcommand{\zO}{\boldsymbol{\hat{O}}}
\newcommand{\zP}{\boldsymbol{\hat{P}}}
\newcommand{\zZ}{\boldsymbol{\hat{Z}}}
\newcommand{\zI}{\boldsymbol{I}}
\newcommand{\zM}{\boldsymbol{M}}
\newcommand{\zn}{\boldsymbol{n}}
\newcommand{\zb}{\boldsymbol{b}}

\newcommand{\circo}{~\raisebox{1pt}{\tikz \draw[line width=0.6pt] circle(1.1pt);}~}

\newcommand{\mymark}[1]{{\color{black}#1}}


\section{Introduction}
A fundamental challenge in photon science is the loss of phase information of the electromagnetic wavefield during measurement. This phase problem plagues the study of light-matter interactions across various energy scales and disciplines, e.g., in radar imaging \cite{Auslander1985,Bonami2007}, astronomy \cite{fienup1987phase, krist1995phase}, microscopy \cite{Kermisch1970,  Shechtman2015, Miao1999, Chapman2006}, and crystallography \cite{Harrison1993}. It also appears in imaging methods using electrons \cite{Coene1992} and neutrons \cite{Allman2000}. It arises because no detector can directly sample the electromagnetic field oscillations of optical and X-ray light. For instance, even the most advanced X-ray detectors, such as micro-channel plates, can only capture the intensity of the wavefield averaged over time windows greater than $10$~ps \cite{Va’vra2009, Tremsin2021}. Meanwhile, reliable algorithms have been developed to retrieve the phase in two dimensions (e.g., diffraction imaging \cite{Fienup1982, Candes2015, Metzler2018}) and higher dimensions (e.g., crystallography \cite{Cowtan2003}). The one-dimensional phase problem is highly ill-posed and inherently more challenging to solve due to multiple non-trivial ambiguities \cite{Beinert2015}. The mathematical proof is derived from D'Alembert's fundamental theorem of algebra, which states that, unlike single-variable polynomials, multidimensional polynomials are generally not factorable \cite{Hayes1982}. Unlike higher-dimensional problems, it is typically not possible to uniquely solve a one-dimensional phase problem using only one measurement, even when prior information such as non-negativity is assumed \cite{Beinert2018}.

One-dimensional phase problems arise, for example, in ultrafast laser pulse diagnostics \cite{Trebino1993, Spangenberg2015}. The laser pulse is only a few femtoseconds long, and its temporal response cannot be measured directly. Instead, the pulse is gated with itself in time with the help of a non-linear optical medium, and its frequency spectrum is measured for different time delays. The temporal shape and length of the pulse are then retrieved from this two dimensional dataset, which is called the frequency-resolved optically gated (FROG) trace \cite{Trebino1997}, using a phase retrieval algorithm based on the short-time Fourier transform \cite{Kishore2016}. Another example is the Griffin-Lim algorithm, which is used to separate speech signals from background noise in two-dimensional audio spectrograms \cite{Griffin1984}.

An analogous problem arises in M\"{o}ssbauer physics, when the nuclear forward scattering (NFS) signal of an object is measured. The recoilless scattering of X-ray photons by nuclei, known as the M\"{o}ssbauer effect \cite{M_ssbauer1958}, offers unique insights \cite{Ko2022, Lohaus2023, Miglierini2012, Knyazev2020} into the magnetic and electronic structure of materials. The sharp natural linewidth of the nuclear transitions allows for extraordinary energy resolutions ($10^{-13}-10^{-8}$ eV) compared to electron spectroscopy methods ($10^{-2}-10^{-1}$~eV) \cite{Fadley2010, Egerton2007}. For example, the 14.4~keV transition in the iron isotope $\Fe$ has an extremely narrow natural linewidth $\Gamma =$ 4.7~neV, corresponding to a quality factor of $\sim10^{12}$. Conventional lab-based methods to measure the energy spectrum of these sharp transitions are unsuitable for materials with unavailable or short-lived radioactive sources \cite{Bessas2014, Knyazev2020}, and for experiments requiring a small, focused beam \cite{Gavriliuk2008, Kothapalli2014}. For the energy-resolved study of $\Fe$-containing materials, synchrotron M\"{o}ssbauer source (SMS) setups \cite{Smirnov1997, Potapkin2012} that use pure nuclear Bragg reflections from a $\Fe\mathrm{B}\mathrm{O}_3$ crystal have been developed. This technique enables $\Fe$ M\"{o}ssbauer spectroscopy at synchrotrons but introduces other challenges. The Doppler motion of the crystal to tune the energy often causes fluctuations in the reflected beam due to crystal imperfections. Moreover, maintaining the temperature stability of the setup is critical for achieving high energy resolution (3-6~$\Gamma$). In addition, high resolution reduces photon flux, resulting in a trade-off between resolution and intensity \cite{Yaroslavtsev2022}.

Instead, the sub-100~ps X-ray pulses from advanced synchrotron sources can be directly used to study the nuclear transitions in time domain. These pulses, with energy bandwidths monochromatized to approximately 1 meV ($\approx 10^4$ times the hyperfine splitting of the resonances), contain fewer than 0.01 resonant photons per pulse. As the synchrotron pulse traverses an object, the entire nuclear ensemble coherently scatters a single resonant X-ray photon, forming a nuclear exciton-polariton \cite{Smirnov1999, Hannon1999, Smirnov2005}. Following this excitation, the exciton undergoes collective evolution and spontaneous decay, resulting in the emission of photons at delayed times.  
The linear response of the object to the weak driving field is described in the energy domain as $\hat{E}_{\text{s}}(\omega) = \hat{O}(\omega)  \hat{E}_{\text{in}}(\omega)$, where $\hat{E}_{\text{in}}$ and $\hat{E}_{\text{s}}$ represent the energy spectra of the input and scattered X-ray fields, respectively, and $\hat{O}(\omega)$ is the transmission function of the object. For X-rays of wavelength $2\pi/k$ passing through an object of thickness $z$, the transmission function is given as follows:
\begin{equation}
    \hat{O}(\omega) = e^{-\mathrm{i}\chi_\mathrm{0} (\omega) kz}.
\end{equation}
It is inherently complex due to the complex susceptibility $\chi_\mathrm{0}$ of the nuclear transition \cite{Roehlsberger2005, Smirnov1999}. We can assume that all spectral components $\hat{E}_\mathrm{in}$ of the input synchrotron pulse have an equal magnitude $E_\mathrm{0}$ within the narrow energy bandwidth of the monochromatizaion. The scattered field $E_{\text{s}}$ is then related to the object's transmission function $\hat{O}$ as
\begin{align}
    E_{\text{s}}(t) \propto \mathcal{F}\{\hat{E}_{\text{s}}(\omega)\}  = E_\mathrm{0} \mathcal{F}\{\hat{O}(\omega)\},
\end{align}
where $\mathcal{F}$ denotes the Fourier transform from the energy to the time domain. In the timing mode of operation, the filling pattern of the electron storage ring is chosen such that synchrotron pulses are temporally spaced at intervals longer than the lifetime of the nuclear transitions. Avalanche photodiodes detect delayed photons as a function of time after excitation, and the measured signal is proportional to the intensity of the scattered field $|E_{\text{s}}(t)|^2$. The hyperfine structure of the object manifests in the beating patterns of this temporal response. 

Despite advances in data analysis software and modeling \cite{Sturhahn2000, nexus2023}, interpreting the time domain response of NFS to extract the different hyperfine parameters remains challenging. On the other hand, if phase information of the photons is available, the inverse Fourier transform can yield the complex energy spectrum of an object from NFS measurements without relying on a fit model or SMS setups. Furthermore, the energy resolution is not limited by the bandwidth of the crystal reflection in the SMS setup. However, the phase shift experienced by the scattered X-ray wavefield is lost in these measurement techniques, presenting a one-dimensional phase problem.

Various methods have been developed to tackle this phase problem in nuclear resonant scattering. For example, interferometry has been attempted to measure the phase shifts of a nuclear forward scattering object using a triple Laue interferometer \cite{Hasegawa1994, Izumi1995}. However, the short wavelengths and near-unity refractive indices of most materials in the X-ray regime make designing and stabilizing such interferometers highly challenging. Contemporary approaches substitute the interferometer with a probe sample mounted on a Doppler drive, where the Doppler drive serves as the phase shifter, and the object and probe samples act as interferometer arms. Techniques such as Heterodyne Phase Reconstruction (HPR) \cite{Callens2005} and Frequency-Frequency Correlation \cite{Wolff2023} are based on this setup, but are only applicable when the nuclear resonances of the probe and the object samples are so detuned in energy that their radiative coupling  \cite{Potzel2001, vanBrck2002, Smirnov2005} can be neglected. Additionally, these methods require a probing beam with a narrow single-energy line, which is scanned in small detuning steps across the object to improve energy resolution. The probe beam can be generated using a nuclear scattering sample that acts as a two-level quantum system. However, to achieve this, it is difficult to eliminate the residual hyperfine interaction in quasi-single-line absorbers such as $\Fe$-based stainless steel \cite{Sahoo2011}.

In this paper, we propose performing ptychography to retrieve the one-dimensional phase of an object. Ptychography is a scanning technique that uses multiple overlapping measurements to constrain the phase problem, and is commonly implemented in the field of X-ray diffraction imaging \cite{Rodenburg2004}. It is a mathematical cousin of the short-time Fourier transform, but has less stringent scanning requirements. For nuclear ptychography, the probe has a broad energy spectrum and can illuminate a wide energy range on the object, allowing the scanning of the object spectrum with fewer measurements. The overlap between the measurements is set by the energy detuning of the probe with respect to the object.

Early conceptual work on a phase-sensitive, ptychographic  method for nuclear resonant systems by Haber \cite{HaberPhD2017}, recognized its potential in the emerging field of hard X-ray quantum optics with nuclear exciton-polaritons, which exhibit long coherence times and unique quantum behaviors \cite{Roehlsberger2012, Roehlsberger2010, Buervenich2006, Haber2016}, and inspire research in fundamental physics \cite{Coussement2002, Liao2015} and quantum information \cite{Liao2022, Velten2024}. Accessing the spectral phase provides insights into the coherence properties of the excitonic state, which is key to its manipulation and control \cite{Tittonen1993, Heeg2015, Bocklage2021}. For example, the phase can help distinguish between incoherent line-broadening due to thickness effects and coherent features due to hyperfine distributions in the transmission spectrum. 

In two-dimensional phase problems, blind ptychography approaches \cite{Thibault2009, Fannjiang2020} are widely used to simultaneously retrieve both the object and probe. We acknowledge the recent publication by Yuan et al. \cite{Yuan2025}, which presents a similar inversion concept for nuclear resonant systems. However, in our experiment, the object transmission spectrum was too sparse (six well-separated $\Fe$ resonances) to support the simultaneous probe retrieval. Instead, we explicitly modeled the hyperfine splittings in the probe spectrum to accommodate them in our reconstruction algorithm. Their probe is a thinner, single-resonance analyzer (1 $\micron$), whereas our use of a thicker 20 $\micron$ stainless-steel foil allows broad spectral coverage and demonstrates experimental robustness to probe-induced thickness effects. Similar to how ptychographic imaging has advanced spatial resolution in diffraction-limited systems \cite{Zheng2013, Jiang2018, Holler2014}, we propose that the development of a ptychographic spectroscopy method could unlock new resolution regimes for M\"{o}ssbauer science.

\begin{figure*}[htbp]
\centering \includegraphics[width=\linewidth]{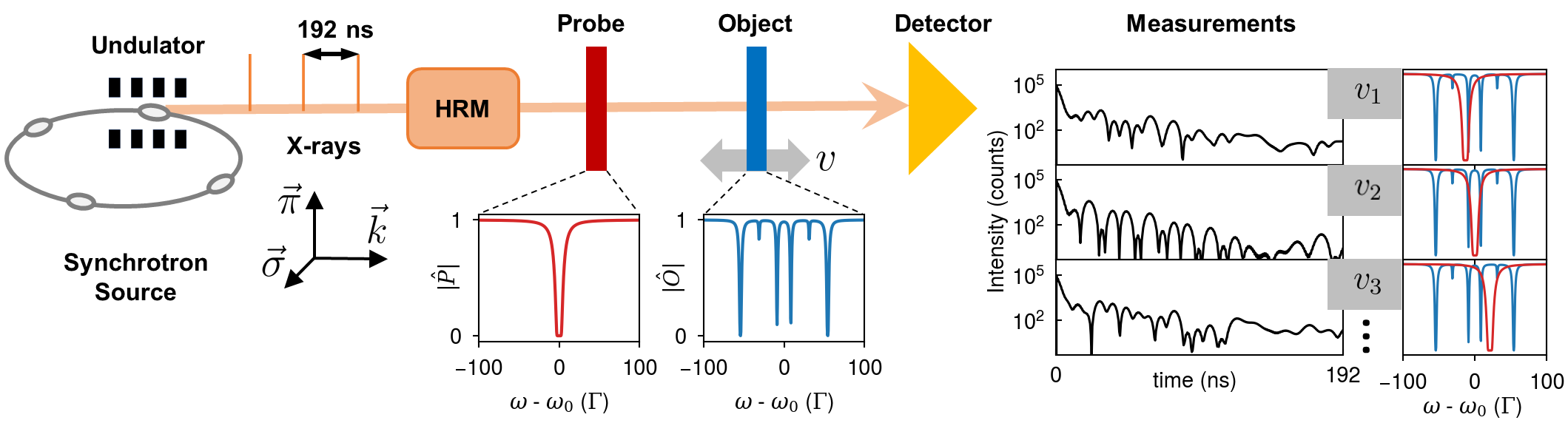}
\caption{\textbf{Ptychography scheme for nuclear forward scattering:} Pulsed X-ray radiation is generated from the synchrotron source with wave vector $\vec{k}$ and linear polarization along $\vec{\sigma}$. It is monochromatized by a high resolution monochromator (HRM) to a bandwidth of 1 meV around the nuclear resonance ($\hbar\omega_0$ = 14.4~keV). A probe sample is mounted in front of an object sample and either of the two is moved with respect to the other with a velocity $\vec{v} \parallel \vec{k}$ using a Doppler drive. The X-ray pulses are scattered by the nuclei in the two samples. The magnitude of the transmission spectrum of the probe $\zP$ and object $\zO$ is shown in the insets. The detector measures the combined response of the samples as counts of photons scattered over time. Changing the velocity of the Doppler drive changes the relative detuning of the samples in the energy domain, and leads to a different temporal response at the detector. This allows multiple intensity measurements (shown for different velocities $v_1, v_2, v_3, \ldots$) to be collected. }\label{fig: exp_setup}
\end{figure*}
\section{Ptychography using resonant scattering by a two-sample system}
The goal of ptychography is to retrieve the complex object transmission function $\hat{O}(\omega)$. To achieve this, a probe sample is placed in front of the object sample in the X-ray beam path (\fig{fig: exp_setup}). In imaging setups, the probing beam is shaped by a lens or aperture that spatially restricts the illumination to a localized spot on the object. In our setup, the probe transmission function $\hat{P}(\omega)$, must have a spectral width comparable to that of the hyperfine splittings of nuclear levels in the X-ray regime. A suitable choice is a quasi-single-line absorber such as stainless steel, whose thickness can be adjusted to achieve the desired spectral width.

Next, a mechanism is required to energetically detune the probe with respect to the object, so that a different energy range in the object spectrum is illuminated for each ptychographic measurement. This can be achieved via the Doppler effect if either the object sample or the probe sample is moved with respect to the other along the direction of beam propagation. This motion induces an additional time-dependent phase shift in the radiation scattered by the probe sample, 
\begin{equation}
    \varphi(D, t) = \dfrac{2 \pi }{ \lambda} |\vec{v}| t = Dt
\end{equation}
where $\lambda$ is the wavelength of the X-rays and $D = 2 \pi |\vec{v}| /\lambda $ is the  Doppler shift (in angular frequency) induced by the relative motion with velocity $\vec{v}$. The combined transmission function of the probe-object sample system is given as
\begin{align} \label{eq: convolve}
    \hat{Z}(D, \omega) 
    &= \hat{P}(\omega + D ) \cdot \hat{O}(\omega).
\end{align}   

 Assuming that all detected photons are coherently scattered, the intensity at the detector at any time is equal to the squared magnitude of the scattered wavefield,  and can be modeled as 
\begin{align} \label{eq: continuous ptychographic model}
    I(D, t) 
               \propto \left| \mathcal{F} \{\hat{P}(\omega + D ) \cdot  \hat{O}(\omega) \}\right| ^2.
\end{align}
This is the one-dimensional continuous ptychographic forward model. It is non-linear due to the presence of the modulus squared $\left| \cdot \right|^2$ operator. Inverting \eq{eq: continuous ptychographic model} to obtain the complex object function $\hat{O} (\omega)$ is impossible using only one measurement. Even by imposing prior constraints on $\hat{O}(\omega)$, such as a compact support or sparsity, the one-dimensional phase problem can have an infinite number of solutions and is unstable \cite{Walther1963, Grohs2020}. 
However, it is possible to use data diversity to impose the overlap constraint \cite{Thibault2009} in \eq{eq: continuous ptychographic model}. Multiple related measurements are taken by changing the detuning $D$ such that the probed parts of the object overlap in energy, as shown in \fig{fig: exp_setup}. This scheme of time and energy-resolved measurement of the scattering process encodes the phase of the one-dimensional object in a two-dimensional ptychographic dataset called a "ptychogram" \cite{GuizarSicairos2021}, and may be recovered using a decoding algorithm. It is analogous to a traditional spectrogram that encodes the variation in a signal's frequency content with time. 

In ptychographic imaging setups, smaller spatial features in a sample result in larger scattering angles in detected diffraction patterns. The largest scattering angle that the detector can capture sets the minimum achievable spatial resolution for the reconstruction. In the nuclear ptychography setup, an analogous constraint arises due to the maximum acquisition time $T_\mathrm{max}$ at the detector, which is determined by the finite time interval between synchrotron pulses. This imposes a limit on the maximum achievable energy resolution of the reconstructed object spectrum (in units of $ \Gamma $)
\begin{equation}\label{eq: det_resolution}
 \hbar\Delta \omega' = \frac{2 \pi}{T_\mathrm{max}}  \cdot 
 \frac{\hbar}{\Gamma} = {2 \pi} \cdot \frac{\tau}{T_\mathrm{max}}
\end{equation}
where $\tau = \hbar/~\Gamma$ is the lifetime of the excited nuclear state. Because nuclear transitions are extremely sharp in energy, their time response can extend beyond $T_\mathrm{max}$, fundamentally limiting the energy resolution of this technique. This contrasts with imaging experiments in which the resolution is typically constrained by factors such as the radiation dose on the sample \cite{Schropp2010}, decoherence effects \cite{Clark2012} and Poisson noise, rather than detector size.

\section{Decoding scheme}\label{sec: decoder}
The ptychogram can be inverted using numerical algorithms, for which we discretely approximate the continuous phase problem in \eq{eq: continuous ptychographic model}. The discrete object and probe functions are expressed as one-dimensional arrays $\zO \in \mathbb{C}^{N}$,  $\zP \in \mathbb{C}^{N }$ on an energy grid of length $N$ and resolution $\Delta \omega$. We take $j = 1,2 \cdots M$ measurements corresponding to different probe detunings $D_j$ and detuned probe functions $\zP_{j, i} = \zP(\omega_i,  D_j)$. The intensity in the time domain is measured and binned into $N$ time points with a fixed interval $\Delta t$ and modeled as $\boldsymbol{{I}}_j \in \mathbb{R}^{N}$. The phase problem can now be formulated as an optimization problem to solve for an object $\zO$ that minimizes a cost function  $\rho : X \rightarrow [0, \infty) $ given by
\begin{align}\label{eq: optim problem}
    \rho (\zO) &=   \sum_{j=1}^{M} \left\Vert \sqrt{\zI_j} - \sqrt{\zb_j} \right\Vert^2\nonumber\\
    &=  \sum_{j=1}^{M}  
    \left\Vert 
    \left|
    \mathbf{F} \left\{ \zP_{j}\circo \zO  \right \}
    \right| 
    - \sqrt{\zb_j}  
    \right\Vert^2.
\end{align} 
where $\mathbf{F}$ represents the discrete Fourier transform, $\circo$ denotes pointwise multiplication and $\|\cdot \|$ denotes the $\ell^{2}$ norm. The cost function represents the distance between the measured intensities $\zb_j$ and the modeled intensities $\zI_j$ of the ptychogram and is based on the Poisson likelihood model for noise in the ptychogram (\cite{Thibault2012}, \mymark{Supplement 1: Sec. S4}). In \eq{eq: optim problem}, we optimize an object of grid size $N \sim 10^3$, where global optimization methods struggle due to the curse of dimensionality \cite{Chen2014, Guirguis2020}. Therefore, a local search with gradient descent is performed to minimize $\rho$ by using its local gradient with respect to the object \cite{Polyak1964}. Owing to the non-convexity of the cost function $\rho$, the gradient descent algorithm may converge to local minima and the uniqueness of the solution is not guaranteed. To mitigate slow convergence, we incorporate a stochastic gradient descent (SGD) algorithm where the ptychogram dataset is shuffled and divided into random ``mini-batches" whose gradients are used to update the object \cite{Monro1951}. In our case, we observe that SGD converges noisier than the classic gradient descent, but it achieves an optimum with an order of magnitude fewer iterations. 

We implemented the reconstruction algorithm for Nuclear Ptychography in a software package which we call NuPty \cite{Nupty2025}. All NuPty algorithms were implemented using PyTorch \cite{Paszke2019}. This enables a flexible and faster analysis of the phase problem because PyTorch uses its automatic differentiation capabilities to efficiently compute gradients and supports GPU-accelerated computations. The NuPty reconstruction scheme takes into account two key nuances of the ptychography experiment:

    \paragraph{1. Multimodal ptychography model:} Thickness variations in the transmitting probe sample may introduce several incoherent scattering paths into the setup.  To account for this, the intensity at the detector is modeled as an incoherent superposition of the intensities of the scattered fields corresponding to different probe modes $m$ illuminating the object \cite{Thibault2013, Li2016}, i.e., in \eq{eq: optim problem}
\begin{align} \label{eq: incoherent forward model}
    \zI_j &= \sum_m  w_m \left|
    \mathbf{F} 
    \left\{ 
     \zP^{(m)}_{j} \circo \zO 
    \right\}
    \right|^2 ,
\end{align}
where $w_m$ is a scalar denoting the relative weight of each probe mode $\zP^{(m)}_{j}$.

\paragraph{ 2. Time window:} Nuclear resonant scattering occurs with a delay (ns timescale) compared with prompt electronic scattering (ps timescale). The time-resolving detector is synchronized to the synchrotron bunch clock and resets to zero when a new X-ray pulse hits the sample. To prevent the prompt signal from saturating the detection system, a veto interval is set around the bunch clock, establishing a data acquisition time window from $T_{\mathrm{min}}$  to $T_{\mathrm{max}}$ for the nuclear scattered signal. To ensure that the reconstruction result is scale-independent with respect to the number of photons $N_\mathrm{ph}$ detected in the time window, the algorithm uses a normalized form of the cost function 
\begin{align}\label{eq: norm optim problem}
    \tilde{\rho} (\zO) &=  \dfrac{1}{ B \cdot f_\mathrm{ph} \cdot L} \sum_{j=1}^{B} \left\Vert \sqrt{\boldsymbol{W}\circo \zI_j} - \sqrt{ \dfrac{ M \cdot \|\mathbf{F}\|^2} {N_\mathrm{ph}/ f_\mathrm{ph}}\cdot \zb_j} \right\Vert^2.
\end{align} 

Here, $B$ is the batch-size, i.e., the number of measurements used at a time to update the gradient, $M$ is the total number of measurements and $L = (T_\mathrm{max} - T_\mathrm{min})/ \Delta t$. To speed up the calculations of the ptychogram and the cost function, we use the fast Fourier transform (FFT) with the operator norm $ \|\mathbf{F}\| = N$. The symbol $f_\mathrm{ph}$ denotes the probability of photon scattering within the time window
\begin{equation}
\boldsymbol{W}(t_i) = 
\begin{cases} 
    1 &  T_\mathrm{min} \leq t_i \leq T_\mathrm{max}, \\
    0 & \text{otherwise}.
\end{cases}
\end{equation}
This probability can be calculated through simulations of the experimental setup and an order-of-magnitude estimate is sufficient for practical purposes.
\begin{figure*}[htbp!]
\centering \includegraphics[width=\linewidth]{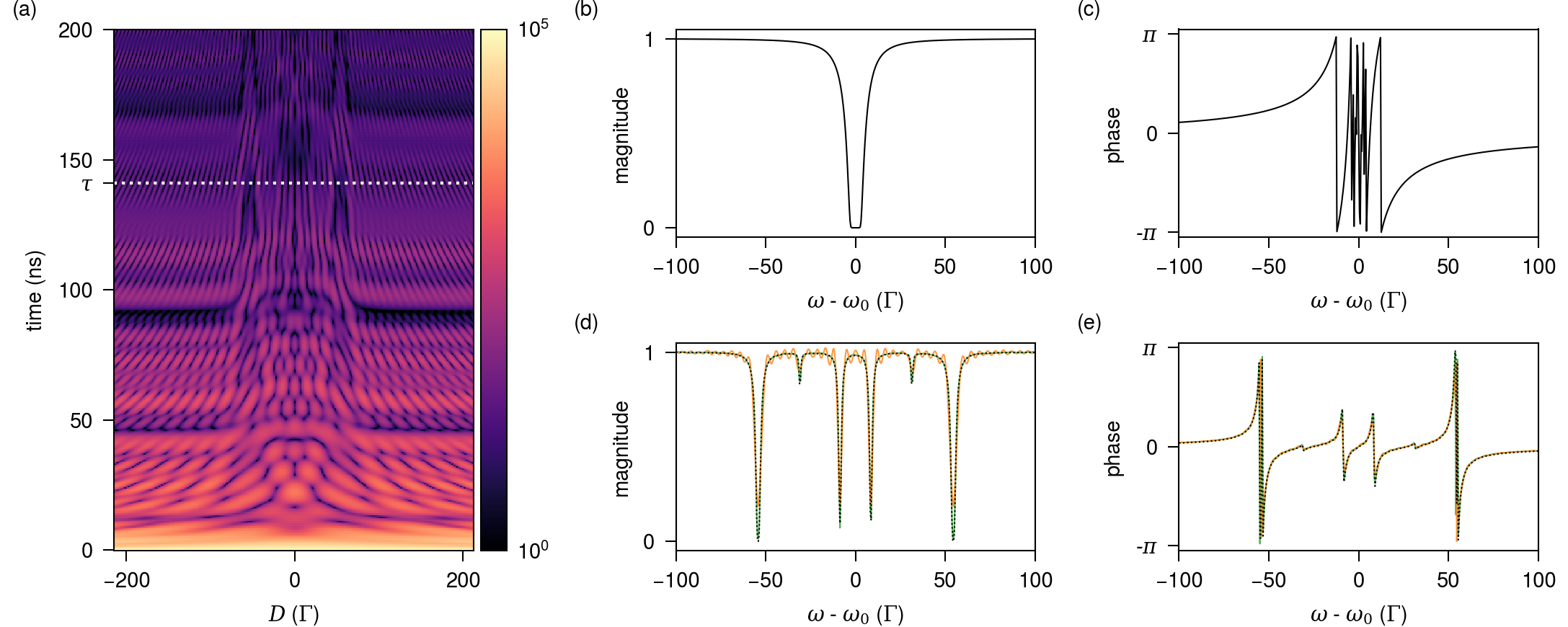}
\caption{\textbf{Phase retrieval from simulation:} (a) Ptychogram simulated as described in Sec. \ref{sec: simulation}. The dotted white line marks $\tau = 141$~ns which is the lifetime of the 14.4 keV energy level of the $\Fe$ nucleus. The magnitude (b) and phase (c) of the probe spectrum are shown. (d) The magnitude of the transmission spectrum of the reconstructed object is plotted for $T_\mathrm{max} =$  $ 200 $~ns (orange) and $T_\mathrm{max} =$ $ 1000 $~ns (green). The reconstructed spectrum in green overlaps perfectly with the true object spectrum (dotted black line). For lower $T_\mathrm{max}$, artificial peaks appear in the reconstructed spectrum.  (e) Reconstructed phase spectrum for the two values of $T_\mathrm{max}$, showing distinct phase shifts at each resonance line. In the plots, $\hbar \omega_0$ refers to the energy of the photons at resonance  = 14.4~keV. }\label{fig: simulated_ptych}
\end{figure*}

\section{Simulation} \label{sec: simulation}
To benchmark the phase retrieval algorithm, a simulation of the experiment was performed using the NEXUS software package \cite{nexus2023} and is shown in \fig{fig: simulated_ptych} (a). A stainless steel foil with an enrichment $95\%$ $\Fe$ and a thickness $20$ $\micron$ was taken as the probe sample. A $95\%$ enriched $\Fe$ metal foil of thickness $2.5$ $\micron $ was taken as the object sample. Figures \fig{fig: simulated_ptych} (b), (c) and (d), (e) represent the simulated energy spectra of the object and the probe samples, respectively. The magnetic structure of the simulated samples is based on the foils used in the real experiment, details of which can be found in \mymark{Supplement 1: Sec. S1-S2}. The probe absorption function has an almost Lorentzian line profile with a full width half maximum (FWHM) of $\sim 10$~$\Gamma$. The broad energy spectrum ensures sufficient overlap between adjacent measurements. The object transmission function is Zeeman split into six lines. The object sample is also modeled after the experimental values (see \sect{sec: results exp}) to contain $94.2(3)\%$ magnetic moments parallel to the $\vec{\sigma}$ direction, which coincides with the linear polarization vector of the incident synchrotron beam, and the remaining $5.8(3)\%$ are isotropically distributed. For both orientations of the magnetic moment, the selection rules of the 14.4~keV $\Fe$ transition prohibit the object sample from being optically active \cite{Roehlsberger2005}. Therefore, we only aim to reconstruct the spectrum of the $\vec{\sigma} \to \vec{\sigma}$ scattering channel. All other entries in the scattering matrix are zero.

To simulate the time-domain scattering signal and perform Fourier transforms efficiently, we used the discrete Fast Fourier Transform (FFT). The number of points $N_\mathrm{FFT}$  used in the Fast Fourier Transform (FFT) of the time-domain signal determines its energy resolution as
\begin{equation}\label{eq: fft_resolution}
    \hbar \Delta \omega  =  \dfrac{2 \pi}{N_\mathrm{FFT} \cdot \Delta t} \cdot \dfrac{\hbar }{~\Gamma} 
\end{equation}
where $\Delta t$ denotes discretization of the temporal grid. To accurately compute the linear convolution of the probe and object signals of length N, $N_\mathrm{FFT} > 2N-1$ to prevent circular convolution errors. A large $N_\mathrm{FFT}$ ensures that $\Delta \omega$ is sufficiently small to resolve the sharp spectral peaks and avoid aliasing errors.

During the experiment, the energy resolution is fundamentally limited by the maximum acquisition time $T_\mathrm{max}$, as shown in \eq{eq: det_resolution}. However, in the simulation, $\hbar \Delta \omega$ can be improved indefinitely by choosing larger values of $N_\mathrm{FFT}$. For $\Delta t = 0.5$~ns, the probe and object spectra are defined in a $N_\mathrm{FFT} = 4096$ point energy grid from $-886.6$~$\Gamma$ to $886.2$~$\Gamma$ in $0.4$~$\Gamma$ steps, where $\Gamma$ = $4.7$~neV. The noiseless ptychogram was simulated by detuning the probe spectrum at $M = 512$ different Doppler shifted energies $\hbar D \in (-200, 200)$~$\Gamma$ with $0.78$~$\Gamma$ steps. Due to the broad probe energy spectrum, this creates an overlap of $\sim 90\%$ between consecutive measurements. 

To evaluate the performance of the phase retrieval algorithm as Poisson noise levels increase, we simulated ptychograms by varying the total delayed photon counts $N_\mathrm{ph}$ and perform the reconstruction (see \mymark{Supplement 1: Sec. S5.A}). We found a roughly linear dependence of the reconstruction accuracy on $N_\mathrm{ph}$, without any anomalies. Thus, the algorithm is stable with respect to increasing levels of Poisson noise. 

To investigate the impact of the time window, the number of photons $ N_\mathrm{ph} $ was fixed at $ 10^9 $ for the next set of tests, whereas only the maximum acquisition time at the detector, $ T_\mathrm{max} $, was varied (see \mymark{Supplement 1: Sec.~S5.B}).
The energy discretization of the FFT grid was kept constant at $\hbar \omega = 0.4$ $\Gamma$, and the time-domain calculations were performed with zero-padding. As shown in \fig{fig: simulated_ptych}(d) and \ref{fig: simulated_ptych}(e), the reconstructed spectra for $T_\mathrm{max} = 200$~ns (orange) and $T_\mathrm{max} = 1000$~ns (green) both show six inverted peaks at the same positions as the simulated spectrum (dotted black line). The X-ray scattering probability depends on the angle between the nuclear magnetic field and X-ray polarization. The higher intensity of the outermost peaks indicates that most of the magnetic moments were aligned parallel to the polarization direction. The outermost peak separation $\Delta E$ can be used to calculate the magnetic hyperfine field 
\begin{equation}\label{eq: B calculation}
    B_\mathrm{hf} = \dfrac{\Delta E}{\mu_\mathrm{N}\cdot\left(3|g_\mathrm{e}| + |g_\mathrm{g}| \right)},
\end{equation}
where $g_\mathrm{g} = 0.18121$ and $g_\mathrm{e} = -0.10348$ are the nuclear $g$-factors for the ground and excited states of $\Fe$, and $\mu_\mathrm{N}$ is the nuclear magneton ($\approx 5.05 \times 10^{-27}~\text{J/T}$). The resulting $B_\mathrm{hf}$ values are 32.73(1) T and 32.72(3) T for the green and orange curves, respectively. For $T_\mathrm{max} = 200$~ns, which is close to the experimental condition of the bunch spacing (= 192~ns), the short time window causes spectral leakage due to sinc interpolation artifacts in the reconstruction. This occurs because zero-padding in the time domain is equivalent to applying a rectangular time window, whose Fourier transform is a Dirichlet kernel. To suppress these artifacts, the measurement time window should extend beyond four nuclear lifetimes between pulses. Additionally, the shortened time window affects both the relative phase shifts and the peak intensities of the resonance spectrum.

\section{Results and discussion}\label{sec: results exp}

\begin{figure*}[htbp!] 
\centering \includegraphics[width=\linewidth]{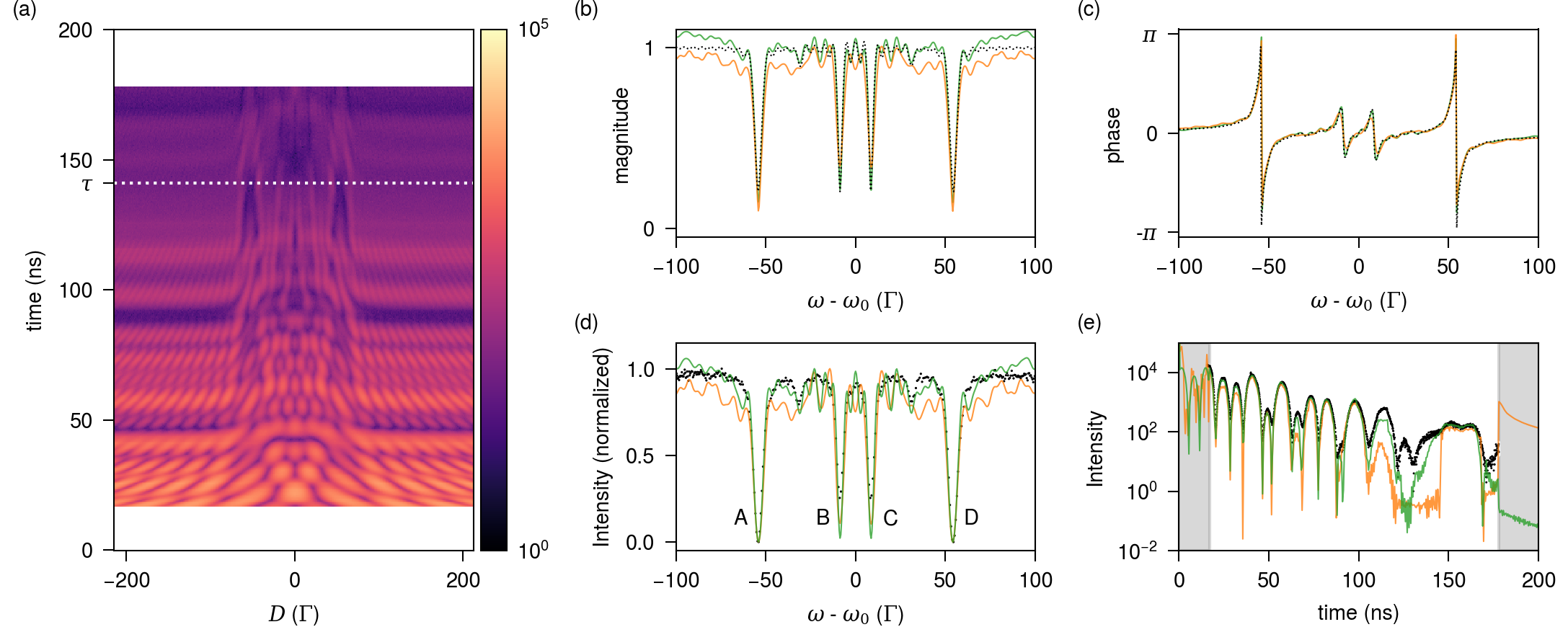}
\caption{\textbf{Phase retrieval from experiment:} (a) Ptychogram measured in the PETRA III experiment. The dotted white line marks $\tau = 141$~ns which is the lifetime of the 14.4 keV energy level of the $\Fe$ nucleus. (b) Magnitude and (c) phase of the reconstructed object spectra for $|\hbar D_\mathrm{max}| = 70$~$\Gamma$ (in green) and $|\hbar D_\mathrm{max}| = 210$~$\Gamma$ (in orange) shown alongside the simulated object spectrum $\zO_H^*$ (dotted black line). (d) Intensity spectrum of the reconstructed objects plotted alongside the measured synchrotron M\"{o}ssbauer spectrum of the $\Fe$ foil used in the experiment. All spectra are normalized from 0 to 1. (e) Time response of the reconstructed objects plotted alongside the measured nuclear forward scattering from the $\Fe$ foil. The gray shaded region lies outside the data acquisition time window of 17 to 178~ns.}
\label{fig: exp_results}
\end{figure*}

We conducted a proof-of-principle experiment at the high-resolution dynamics beamline P01 at the PETRA III synchrotron source, providing X-ray pulses at 192~ns intervals in the timing mode of operation. X-rays were monochromatized to one meV bandwidth at the 14.4~keV $\Fe$ nuclear transition. The probe and object foils were characterized using NFS measurements. 
The probe is given by a stainless steel foil, enriched to $95\%$ in $\Fe$, with a Lamb-Mössbauer factor of 0.78. The thickness of the foil approximately follows a normal distribution, centered at 17.8 $\micron$ with a FWHM of 0.7 $\micron$. The object under study is an iron metal foil, enriched to $95\%$ in $\Fe$, with a Lamb–Mössbauer factor of 0.80 and a mean thickness of 2.4 $\micron$ (FWHM 0.3 $\micron$).

The object foil was mounted on a Doppler drive and moved relative to the probe foil with a sinusoidally changing velocity profile. The drive velocity was tuned to a maximum of $20.7$~mm~$\mathrm{s}^{-1}$, to ensure Doppler detunings in the range $\hbar D \in (-210,210)$~$\Gamma$. A 0.12~T magnetic field ($\vec{B} \parallel \vec{\sigma}$, beam polarization) was applied to the object. Photons were detected using silicon-based avalanche photodiodes with a time resolution of $\sim 0.3$~ns and binned into time channels such that $\Delta t =0.5$~ns. Their arrival times was recorded together with the Doppler velocity at the moment of detection using a multichannel data acquisition system. More details on the samples and setup are available in \mymark{Supplement 1: Sec.~S1-S2.}

Phase information was captured in the ptychogram (\fig{fig: exp_results}(a)), with time and Doppler detuning as its axes. The measured data closely resemble the simulated ptychogram (\fig{fig: simulated_ptych}(a)), except for the experimental data acquisition window from $T_{\mathrm{min}} = 17$~ns to $T_{\mathrm{max}} = 178$~ns. This is due to the vetoing of the electronic scattering signal as described in Sec. \ref{sec: decoder}. The results of the ptychographic reconstruction using the experimental measurements are also shown in \fig{fig: exp_results}. According to our simulations, the majority of the incident photons undergo prompt electronic scattering within the first few picoseconds. Only a small fraction, $ f_\mathrm{ph} \approx 0.019$, is scattered within the delayed time window, resulting in the detection of approximately $ 2 \times 10^8 $ total delayed photons. While reconstructing the object, the algorithm can reasonably extrapolate the missing intensities between 0 and 17~ns by taking advantage of the oversampled measurements. This is also performed in conventional two-dimensional ptychographic imaging in the presence of a beam stop \cite{ Dejkameh2024, Reinhardt2017}. However, there is no information in the ptychogram for times beyond $T_\mathrm{max}$, where extrapolation obviously does not work. For a maximum acquisition time $T_\mathrm{max} = 178$~ns at the detector, the energies of the nuclear transitions are convolved with a sinc function with a main lobe width
of $2 \hbar \Delta \omega' \approx 9.3$~$\Gamma$, which is roughly 23 times larger than $ \hbar \Delta \omega$. To evaluate the quality of the reconstruction despite the finite $T_\mathrm{max}$, we define a filtering window $\boldsymbol{H}$ which is a discretized Heaviside step function:
\begin{equation}
\boldsymbol{H}(t_i) = 
\begin{cases} 
    1 & t_i \leq T_\mathrm{max}, \\
    0 & \text{otherwise}.
\end{cases}
\end{equation}
Applying this filter to the simulated complex transmission spectrum of the object $\zO^{*}$ (from \sect{sec: simulation}) yields
\begin{equation}
\zO^{*}_H = \mathbf{F}^{-1} \{ \boldsymbol{H} \circo \mathbf{F} \{ \zO^{*}  \}
\},
\end{equation}
where $\mathbf{F}$ is the FFT. The resulting transmission spectrum, $\zO^{*}_H$, accounts for sinc artifacts similar to those caused by the experimental measurement window. In \fig{fig: exp_results} (b) and \ref{fig: exp_results}(c), the reconstructed object is compared with this filtered object.

The reconstruction of the complex transmission spectrum from the full data range $ \hbar D \in (-210, 210) $$~\Gamma $ (orange) deviates from the true object, while the limited range $ \hbar D \in (-70, 70) $$~\Gamma $ (green) achieves a closer match. We attribute this behavior to the coupling between the resonance peaks of the probe and the object in this regime (\mymark{see Supplement 1: Sec.~S5.C}). This contrasts to the interference signal used in other techniques  \cite{Callens2005, Wolff2023} where the probe and object spectra are so detuned that the scattered field at the detector can be approximated as $ Z(D, t) = \mathcal{F} \{ \hat{Z}(D, \omega) \} \approx  P(D, t) + O(t)$, where $P(D,t)$ and $O(t)$ are their respective temporal responses. The coupling signal has a higher information density and is less susceptible to background noise, velocity drive calibration errors and incoherent contributions to the data due to the thickness variations in the samples. 

The reconstructed phase enables the calculation of an energy-domain spectrum for the object which can then be compared to its measured synchrotron M\"{o}ssbauer source (SMS) spectrum. The positions of the four most prominent Lorentzian lines in \fig{fig: exp_results}(d) are fitted using a least squares fitting algorithm and are listed in Table \ref{tab: peak_pos}. The four peak positions extracted from the reconstructed object in green match the SMS spectrum up to $\pm $ 0.2~$\Gamma$, which is comparable to the resolution of the calculation grid (0.4~$\Gamma$). According to \eq{eq: B calculation}, the magnetic hyperfine field extracted from the reconstructed outer peak positions is 32.62(8) T, compared to 32.73(4) T from the SMS spectrum. To calculate the preferential orientation of the magnetic domains in the $\Fe$ foil, other peak properties, such as their relative heights and symmetry, are needed. Our reconstructed transmission spectrum reveals the presence of a dominant magnetic moment component parallel to the X-ray polarization direction. In \fig{fig: exp_results}(d), two additional small resonance peaks appear in the SMS spectrum at roughly $\pm$ 31.5~$\Gamma$, due to a minor isotropic nuclear spin component of $\Fe$ ($\approx 5.8(3)\%$ from nuclear forward scattering measurement). Although these smaller peaks cannot be distinguished clearly from the artificial peaks in the reconstructed spectrum, presence of their subtle signatures indicates that nuclear ptychography is already approaching the sensitivity required to detect such fine features, implying a strong potential for further improvements in reconstruction accuracy with longer measurement time windows.

In \fig{fig: exp_results}(e), the time domain response of the reconstructed object shows that the algorithm accurately reconstructs the measured intensities, except for times between 120 and 140~ns. In this region, "bunch addition" incoherence in the experimental data is strong due to the finite time gap of 192~ns between the incident X-ray pulses \mymark{(see Supplement 1: Sec.~S3.B)}. The detector cannot distinguish between photons arriving at $t > 192$~ns after the incidence of the current pulse and those arriving from the next pulse at $t - 192$~ns, causing systematic errors in the measured data. This incoherent contribution to the data cannot be taken into account by the ptychography algorithm while solving the phase problem and therefore affects the reconstruction.

\begin{table}[htbp]
\begin{center}
\caption{Positions of the peaks marked in \fig{fig: exp_results}(d).}
\label{tab: peak_pos}
\resizebox{0.75\linewidth}{!}{ 
\begin{tabular}{ l | c | c | c| c}
 \textbf{Peak} & A & B & C & D \\
 \hline
 \textbf{SMS spectrum} & -54.21(2)  $\Gamma$ & -8.64(3)  $\Gamma$& 8.51(3) $\Gamma$& 54.4(1) $\Gamma$\\
 \hline
 \textbf{Reconstruction (green) } & -54.06(3)  $\Gamma$& -8.54(4)  $\Gamma$& 8.59(3) $\Gamma$& 54.2(3)  $\Gamma$\\ 
\end{tabular}
} 
 \end{center}
\end{table}

The transverse coherence length of the setup is only a few nanometers (\mymark{see Supplement 1: Sec.~S3.A}). The NFS fit on the stainless steel probe foil predicts a root-mean-square surface thickness variation of approximately $0.7$ $\micron$. Therefore, to reconstruct the object spectrum, the multimodal ptychography model was used, where the complex transmission spectrum of the probe sample was simulated at eleven distinct points sampled from its thickness distribution.

\begin{figure}[htbp] 
\centering
\includegraphics{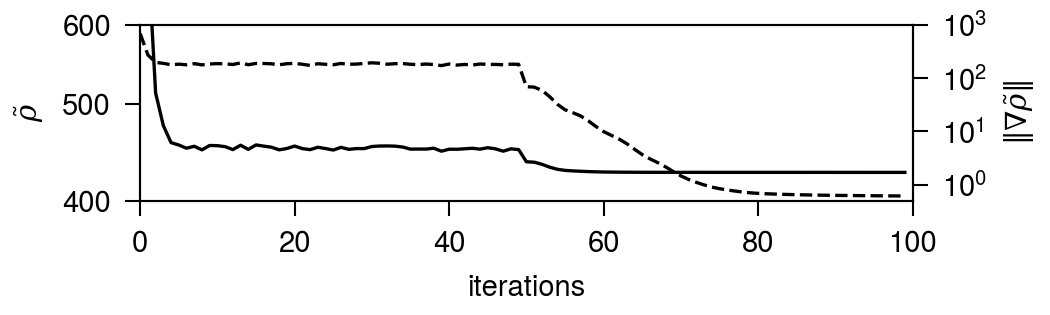}
 \caption{\textbf{Convergence of the phase retrieval algorithm:} Reconstruction cost $ \tilde{\rho} $ (solid line) and the norm of its gradient $ \|\nabla \tilde{\rho}\| $ (dashed line) as the iterations increase. The algorithm converges to a solution in 100 iterations for the reconstruction from experimental data with $|\hbar D_\mathrm{max}| = 70$ $\Gamma$ (shown in green in \fig{fig: exp_results}(b)-(e)).
}\label{fig: iterations}
\end{figure}

The starting object guess was taken as cells with entry 'one'. As shown in \fig{fig: iterations}, the algorithm converges to a solution in approximately 100 iterations. The first 50 iterations use stochastic gradient descent with a batch size of $ B = 20 $, leading to rapid improvement within the initial 10 iterations, followed by stagnation to a local minimum. Due to stochastic noise, the gradient no longer decreases significantly. To refine the solution, the remaining 50 iterations employ standard gradient descent ($ B = M $), reducing stochastic noise and enabling further optimization.

\section{Conclusion and outlook}

We have demonstrated that ptychography provides a powerful framework to solve a one-dimensional phase problem and mitigate its instabilities by redundantly capturing overlapping information in a two-dimensional dataset. We used the method to reconstruct the energy resolved complex spectrum of a magnetized $\Fe$ foil and calculate its hyperfine parameters. The retrieved values are in good agreement with the results predicted by the established measurement techniques of Synchrotron M\"{o}ssbauer source (SMS) spectroscopy and nuclear forward scattering.

Our analysis shows that the energy resolution of nuclear ptychography is fundamentally related to the length of the temporal detection window. Quantum beats between closely spaced nuclear energy levels, split by less than $1 ~\Gamma $, interfere on timescales longer than $\tau = {\hbar}/{\Gamma}$, the natural lifetime of the Mössbauer nucleus. For detection windows extending beyond $T_\mathrm{max} > 2 \pi \hbar/\Gamma$, energy resolution of sub-1~$\Gamma$ becomes achievable (see \eq{eq: det_resolution}). Our method can thus surpass the limits set by the linewidth of the lab source of the gamma ray or the SMS crystal. However, to realize this improved resolution,  increased pulse spacing between X-ray bunches is required. This, in turn, demands low bunch synchrotron timing modes which reduce the overall beam current and brightness. In our experiments, the 192~ns X-ray pulse spacing - although adequate - was not very long compared to the lifetime of $\Fe$, resulting in reconstruction artifacts. Other Mössbauer isotopes, such as $\Sn$ and $\Eu$, have significantly shorter lifetimes and would benefit from the high resolution offered by nuclear ptychography, especially since no suitable SMS is available for them. The synchrotron timing mode can be optimized to balance the X-ray brilliance and pulse intervals and enable the measurement of delayed responses over multiple lifetimes of these M\"{o}ssbauer nuclei.

In conclusion, we have shown that one-dimensional nuclear ptychography provides a robust and versatile tool for exploring complex nuclear resonant phenomena. We can retrieve the spectral phase of the nuclear resonant scattering, which provides direct insight into how an X-ray pulse is reshaped in time as it passes through the nuclear system. This phase information is essential for understanding and engineering coherent phenomena such as electromagnetically induced transparency-like behavior in multilayered X-ray cavities with M\"{o}ssbauer nuclei \cite{Roehlsberger2012, Heeg2013}, where sharp energy-dependent phase shifts from multiple scattering modulate the effective driving field on the nuclei. Beyond solving the phase problem, nuclear ptychography can enable sub-linewidth energy resolution and spectroscopic investigation of M\"{o}ssbauer isotopes at synchrotrons, offering a pathway to resolve elusive features in nuclear spectra, such as the debated existence of magnetic order in hcp-iron~\cite{Bessas2020,Taylor1982}. Our results pave the way for advanced implementations of this method in grazing-incidence and polarization-resolved geometries,  enabling the study of nanostructured and anisotropic systems. The extension of the technique to X-ray free electron lasers offers exciting opportunities to study non-linear effects on the phase of the scattering. Together, these capabilities represent a significant step towards a new era in nuclear quantum optics.

\vspace{ 2 cm}

\begin{center}
\textbf{\Large Supplement 1: Energy-Time Ptychography for one-dimensional Phase Retrieval}
\end{center}
\setcounter{section}{0}
\setcounter{equation}{0}
\setcounter{figure}{0}
\setcounter{table}{0}
\setcounter{page}{1}
\makeatletter
\renewcommand{\theequation}{S\arabic{equation}}
\renewcommand{\thefigure}{S\arabic{figure}}
\renewcommand{\thetable}{S\arabic{table}}
\renewcommand{\thesubsection}{\Alph{subsection}.}

\newcommand{\SSS}{{}^{57}\text{SS}}
\newcommand{\alg}[1]{\textbf{\ref{#1}}}
\newcommand{\imag}{\mathrm{i}}
\newcommand{\vvv}{ v }
\newcommand{\zB}{\boldsymbol{{B}}}


\section{Characterization of the Samples}

To benchmark the ptychography algorithm and perform phase retrieval, the complex energy response of the probe and the object foils is required. Therefore, we characterize both samples mentioned in the main text to extract their structural and hyperfine parameters.

\subsection{Stainless Steel Foil}\label{sec: thick ss foil}

The time response of the stainless steel foil was measured at PETRA III, P01. It was fitted using differential evolution \cite{Biscani2021} followed by Levenberg-Marquardt algorithm in NEXUS \cite{nexus2023}, with a correction for bunch spacing (192 ns). The foil has a Lamb-M\"ossbauer factor of 0.78 and a density of 7.80 g $\text{cm}^{-3}$. In the austenitic phase, stainless steel does not exhibit ferromagnetic order, and the $\Fe$ nuclei exhibit a single-line resonance at 14.4 keV with no Zeeman splitting.

\begin{figure*}[htbp]
\centering \includegraphics{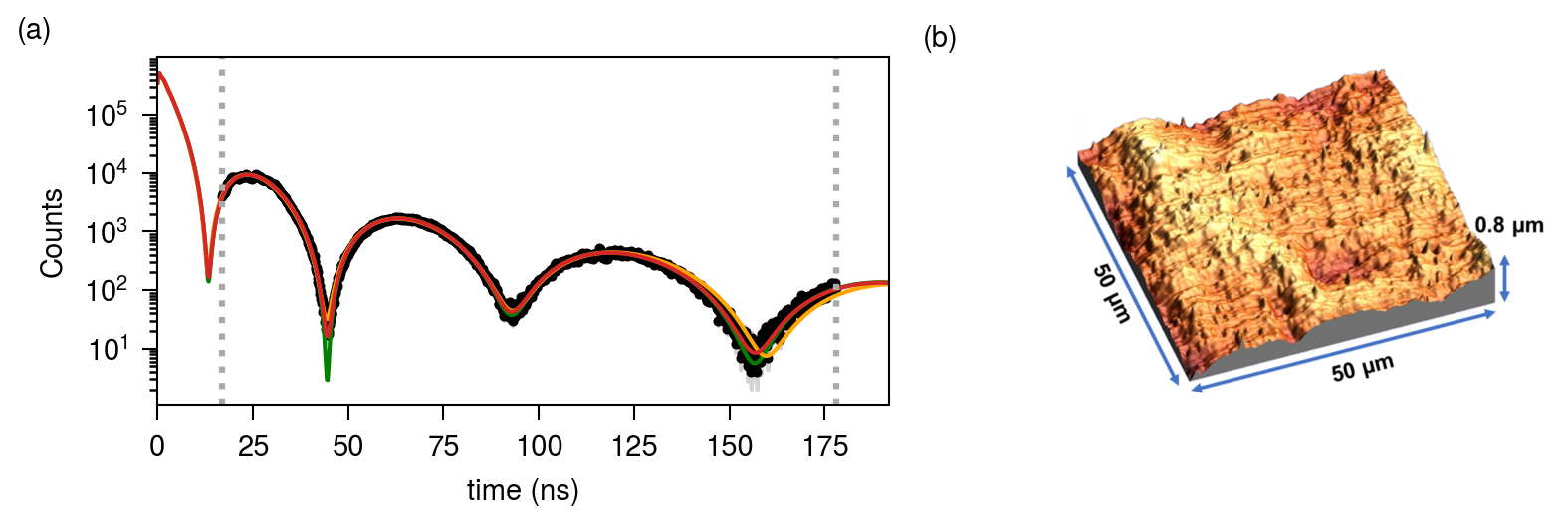}
\caption{\textbf{Characterization of the stainless steel foil:} 
(a) Measured time response (black) of the foil between 17 and 178 ns, shown alongside three fits. The green curve assumes only isomer shift variation, the orange curve assumes only thickness variation, and the red curve represents the best fit using a mixture of both mechanisms. 
(b) Atomic force micrograph showing topographic variation of the foil surface over an area comparable to the beam spot. Similar micrographs were taken over different spots on both sides of the foil, roughly at its center. }
\label{fig: TS_fit_SS}
\end{figure*}

The time response, shown in \fig{fig: TS_fit_SS}(a), exhibits clear dynamical beats with minima around 45, 95, and 155 ns. Several models were tested to explain the line broadening in the energy response of the foil. A model assuming purely coherent broadening due to an isomer shift distribution fails to reproduce the minima in the time response at longer times. A model based only on incoherent broadening caused by foil thickness variation provides a better fit in the intermediate time range but deviates after 125 ns. The best-fitting model includes a combination of both broadening mechanisms: coherent broadening from a distribution of local isomer shifts (due to variations in the electronic environment of the nuclei) and incoherent broadening from thickness inhomogeneity.

The thickness variation is modeled as a Gaussian distribution centered at 17.83(1)~$\mu$m with a full width at half maximum (FWHM) of 0.78(4)~$\mu$m. This variation is consistent with measurements from atomic force microscopy, which show surface height variations between 0.75 and 1.20~$\mu$m across different regions of the foil surface, as shown in \fig{fig: TS_fit_SS}(b). The parameters corresponding to the best fit (shown in red in the figure) are summarized in Table~\ref{tab: TS_fit_SS_foil}. The resulting complex transmission function, simulated using these fitted parameters, is shown in Fig.~5 in the main text.

\begin{table}[htbp]
\begin{center} 
\caption{Fitted parameters of the time response of the thick $\SSS$ foil.}
 \label{tab: TS_fit_SS_foil}
\begin{tabular}{l|c} 
 \textbf{Parameter} & \textbf{Fit Result} \\ \hline  
 Thickness             & 17.83(1) $\mu$m \\ \hline 
 FWHM Thickness         & 0.78(4) $\mu$m \\ \hline 
 FWHM Isomer Shift      & 0.23(1) mm $\mathrm{s}^{-1}$ \\ 
\end{tabular}
 \end{center}
\end{table}

\subsection{Iron Foil}\label{sec: Fe foil}
\begin{figure*}[htbp]
\centering \includegraphics{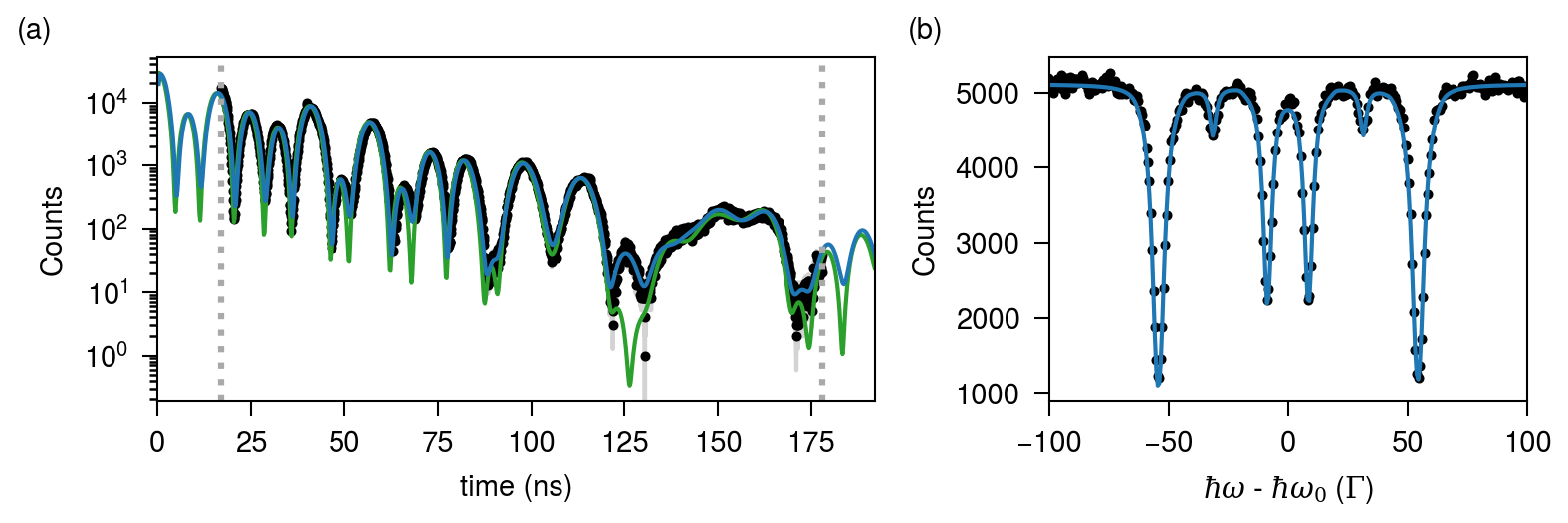}
\caption{\textbf{Characterization of the iron foil:} (a) Measured time response (black points) vs. fitted models with (blue line) and without (green line) bunch spacing correction. (b) Measured synchrotron M\"{o}ssbauer spectrum of the foil (black points) plotted against the fitted model (blue line).}
\label{fig: TS_fit_Fe}

\end{figure*}

The iron foil is 95\% enriched with the resonant $\Fe$ isotope and was subjected to a 0.12~T magnetic field along the electric vector of the polarization of the synchrotron beam during measurements. To characterize the hyperfine parameters, we measured the time response of nuclear forward scattering at PETRA  III, P01 and the synchrotron M\"{o}ssbauer spectrum at at the beamline ID14 (formerly ID18) at ESRF, Grenoble, France. They were fitted simultaneously using a differential evolution and Levenberg Marquardt algorithm in NEXUS. The foil's Lamb-M\"{o}ssbauer factor and density are taken as 0.80 and 7.87~g~$\text{cm}^{-3}$, respectively. We model the foil with two hyperfine sites: site 1 is magnetized along the beam polarization, while site 2 has no preferred magnetic orientation. Line broadening is modeled with a distribution in the thickness of the foil. The best fit (indicated in blue color in \fig{fig: TS_fit_Fe}) was obtained after correcting for bunch spacing (192 ns) from the synchrotron source (See Sec. \ref{sec: incoherent effects}.B). The fit gives a thickness variation in the foil with mean 2.47(1)~$\mu$m and FWHM 0.52(6)~$\mu$m, with 94.2(3)\% of $\Fe$ nuclei magnetized along the  direction of the external magnetic field. The remaining percentage is attributed to misaligned magnetic domains in the foil, possibly due to inhomogeneities or internal strain. The hyperfine magnetic field at both sites is approximately 32.7 T. The list of fit parameters is given in Table \ref{tab: TS_fit_Fe_foil}.

\begin{table}[htbp!]
\begin{center} 
\caption{Fit parameters of the time response of the $\Fe$ foil, shown in \fig{fig: TS_fit_Fe}.}
\label{tab: TS_fit_Fe_foil}
\begin{tabular}{l|c}
 \textbf{Parameter} & \textbf{Fit Result} \\ 
 \hline
 Thickness     &  2.47(1) $\mu$m\\
 FWHM Thickness & 0.52(6) $\mu$m\\
 \hline
 Site 1: Weight  & 94.2(3) \% \\
 Site 1: Magnetic Field & 32.68(1) T \\
 \hline
 Site 2: Weight  & 5.8(3) \% \\
 Site 2: Magnetic Field & 32.7(1) T \\
\end{tabular}
\end{center}
\end{table}


\section{ Details of the experiment}

\begin{figure}[htbp!]
\centering
\includegraphics{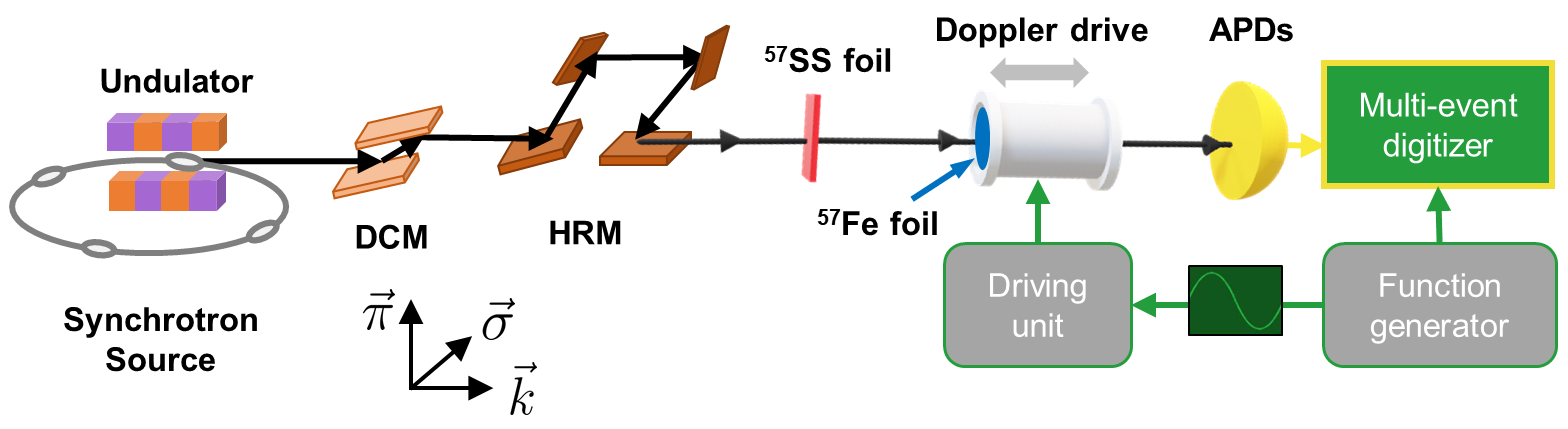}
\captionof{figure}{\textbf{The ptychogram measurement setup at P01, PETRA III:} X-rays from the synchrotron are monochromatized around the $\Fe$ nuclear resonance by the double crystal monochromator (DCM) and high resolution monochromator (HRM).  The object is mounted on a Doppler drive and moved along the direction of the probing beam by the driving unit. A digital function generator is used to give a sinusoidal velocity profile to the drive. The MCS6A collects the photon counts with their time of arrival at the APDs and the instantaneous velocity of the drive.}
\label{fig: ptycho layout}
\end{figure}

The experiment was conducted at beamline P01, PETRA III, using the setup shown in \fig{fig: ptycho layout}. The synchrotron storage ring was operating in the 40-bunch mode with X-ray pulses spaced by 192 ns (pulse duration ~100 ps, repetition rate 5.205 MHz). The X-rays were monochromatized to a bandwidth of ~1 meV around the 14.41 keV nuclear transition of \(^{57}\text{Fe}\) using a silicon double-crystal monochromator (DCM) and a silicon high-resolution monochromator (HRM). The beam cross section on the samples was 18 $\mu$m $\times$ 50 $\mu$m. Detection was performed using a silicon-based avalanche photodiode (APD) array with dark count rates in the tens of mHz, ensuring photon detection dominated by Poisson noise from the X-rays. Each APD is sized ~1 cm × 1 cm and can handle photon fluxes up to \(10^7 \, \text{ph/s}\), with photon-absorbing foils used for protection at higher flux levels. 

The object foil, mounted on a Doppler drive (Wissel MVT-1000), was sinusoidally moved using a driving unit (MR-360). The frequency of the drive was tuned to $\sim 27$ Hz, which is close to its natural resonance frequency ($\sim 25$ Hz), with minimal velocity error $(0.1\%-0.5\%)$. This is equivalent to taking "fly scans" \cite{Huang2015} in ptychographic imaging. The drive achieved a maximum velocity of ~20.7 mm/s, resulting in a Doppler detuning range \(\hbar D \in (-210, 210) \, \Gamma\) and maintaining $>90\%$ measurement overlap to ensure algorithmic convergence. A digital function generator (DFG-1000) controlled the drive signal. An external magnetic field (\(|\vec{B}| = 0.12 \, \text{T}\)) was applied to the object parallel to the electric vector of the beam's polarization (\(\vec{B} \parallel \vec{\sigma}\)).

Photon detection was synchronized via a multi-event digitizer (FAST ComTec MCS6A), which recorded not only the photon arrival times from the APDs, but also the drive velocity from the function generator sampled into 1024 equidistant channels. The MCS6A labels photons with their times of arrival and velocity channel numbers. The data is then histogrammed into a two-dimensional ptychogram. The MCS6A time resolution is 0.1 ns, but the counts were binned into channels separated by 0.5 ns.

To retrieve the phase from the ptychograms, we need the probe detunings relative to the object, which is given by the Doppler drive velocities. Before the experiment, the drive velocity at each MCS6A channel is calibrated using a known probe and object. The detuning for the \(j\)-th channel is given by:
\[
D_j = \frac{\omega_0}{c \Gamma}  \left( \vvv_{\text{max}} \cdot \cos \left( 2 \pi \cdot \frac{j - j_0}{1024} \right) + \vvv_{\text{iso}} \right),
\]
where \(\hbar \omega_0 = 14.41 \, \text{keV}\) and \(\Gamma = 4.66 \, \text{neV}\) for \(\Fe\), \(\vvv_{\text{max}}\) is the maximum velocity, and \(j_0\) corresponds to the channel with maximum velocity. We have also included a term $\vvv_\mathrm{iso}$ to capture the isomer shift between the two samples. The calibration is done by simulating the ptychogram using the detunings with Poisson sampling such that we get the same number of total counts as the experiment. The velocity parameters are then fitted via least squares. The process is outlined in Algorithm \ref{algo: v_calibration}, where the simulated photon counts are compared to the measured counts. The algorithm starts with initial guesses for \(\vvv_{\text{max}}\) and \(\vvv_{\text{iso}}\), and iterates until the best-fit values are obtained.
To start the algorithm, we initialized $(\vvv_\mathrm{max},\vvv_\mathrm{iso}, j_0)$= (25 mm s$^{-1}$, 0 mm s$^{-1}$, 1).
Using the parameters of the pre-characterized stainless steel and \(\Fe\) foils for simulations, we find \(\vvv_{\text{max}} = 20.70(3)\, \text{mm s$^{-1}$}\) and \(\vvv_{\text{iso}} = -0.10(1)\, \text{mm s$^{-1}$}\).

\begin{algorithm}
\caption{Velocity Calibration}
\label{algo: v_calibration}
\begin{algorithmic}[1]
\State \textbf{Input:} Photon counts $\{c_0, c_1, \dots, c_{M-1}\}$, energy domain responses $\zO^*$, $\zP^*$, initial guesses for maximum velocity $\vvv_\mathrm{max}$, isomer shift $\vvv_\mathrm{iso}$, and channel $j_0$
\State \textbf{Output:} Calibrated $\vvv_\mathrm{max}$, $\vvv_\mathrm{iso}$

\State \textbf{Step 1:} Calculate detunings $D_0, D_1, \dots, D_{M-1}$ using current guesses for $\vvv_\mathrm{max}$, $\vvv_\mathrm{iso}$, and $j_0$:\\
 \hspace{1cm} $D_0, D_1, \dots, D_{M-1} \gets \mathrm{CALC\_DETUNINGS}(\vvv_\mathrm{max}, \vvv_\mathrm{iso}, j_0)$

\State \textbf{Step 2:} Simulate the ptychogram using the calculated detunings and object/probe responses:\\
 \hspace{1cm} $\{\zb'_0, \zb'_1, \dots, \zb'_{M-1}\} \gets \mathrm{SIM\_HISTOGRAM}(\zO^*,\zP^*, D_0, D_1, \dots, D_{M-1})$

\State \textbf{Step 3:} Calculate the photon counts from the simulated ptychogram:\\
 \hspace{1cm} $c'_0, c'_1, \dots, c'_{M-1} \gets \sum_{q} \zb'_0, \sum_{q} \zb'_1, \dots, \sum_{q} \zb'_{M-1}$

\State \textbf{Step 4:} Perform least squares fitting to update $\vvv_\mathrm{max}$ and $\vvv_\mathrm{iso}$:\\
\hspace{1cm} $(\vvv'_\mathrm{max}, \vvv'_\mathrm{iso}) \gets \mathrm{LSQFIT}(\{c_0, c_1, \dots, c_{M-1}\}, \{c'_0, c'_1, \dots, c'_{M-1}\}, \vvv_\mathrm{max}, \vvv_\mathrm{iso})$

\State \textbf{Step 5:} Update the velocity parameters:\\
 \hspace{1cm} $\vvv_\mathrm{max} \gets \vvv'_\mathrm{max}, \vvv_\mathrm{iso} \gets \vvv'_\mathrm{iso}$

\State \textbf{Step 6:} Return calibrated $\vvv_\mathrm{max}$, $\vvv_\mathrm{iso}$

\end{algorithmic}
\end{algorithm}

For the phase reconstruction, a stepsize of 0.001 and momentum 0.66 were chosen to run the NuPty gradient descent algorithm.


\section{Incoherent contributions to experimental data} \label{sec: incoherent effects}
\subsection{Thickness variations in the probe}
\begin{figure}[htbp!]
\centering
  \includegraphics{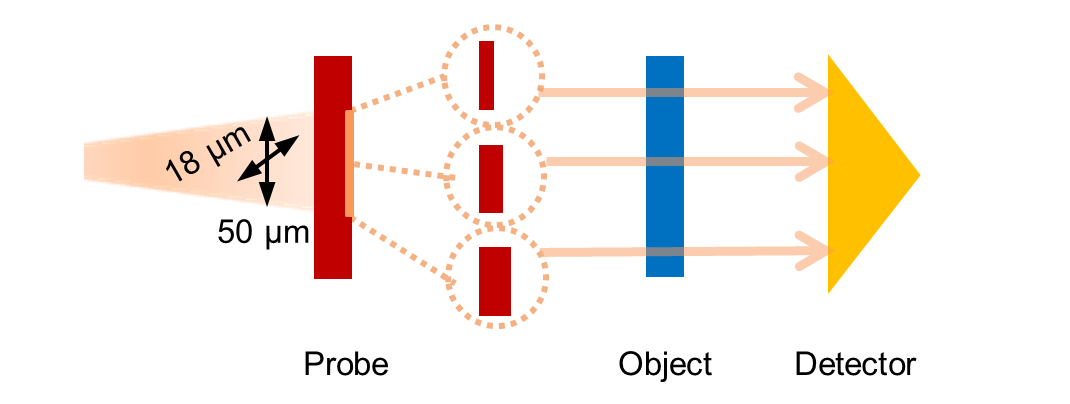}
  \caption{\textbf{Incoherence effects due to thickness variations}: The synchrotron beam has a spot size of $18$ $\mu\text{m}$ $\times$ $50$  $\mu\text{m}$ on the probe. Varying thicknesses of the probe lead to several beam paths traversing the object, which sum up incoherently at the detector.}
  \label{fig: coherence setup}
\end{figure}

Transverse coherence refers to the length scale perpendicular to the beam where phase correlations in the wave field remain stable at a given time. Assuming the synchrotron beam is monochromatic with wavelength $\lambda$, angular size $\sigma_0$ and upstream distance $S$, the transverse coherence length $\xi_T$ is usually approximated as
\begin{equation}
    \xi_T \sim \dfrac{\lambda S}{2 \pi \sigma_0}.
\end{equation}

However, in our experiment, the detector size also significantly affects $\xi_T$, as avalanche photodiodes integrate all the photons hitting their surface. We can calculate this by considering the time domain analogue of the double slit experiment, as explained in Ref. \cite{Baron1999}. The effective transverse coherence length is given as 
\begin{equation}
    \xi_{Te} = \dfrac{\lambda}{2 \pi } \dfrac{1}{\sigma_{Te}},    
\end{equation}
 where $\sigma_{Te} = \sqrt{ \left(\dfrac{\sigma_0}{S} \right)^2 + \left(  \dfrac{\sigma_L}{L}\right)^2}$. Here, $S$ is the distance from the source to the sample, and $L$ is the distance from the sample to the detector. The symbols $\sigma_0$ and $\sigma_L$ represent the vertical sizes of the source and the detector, respectively.
For our experiment, $\lambda = 0.086 $ nm, $L \sim 1$ m and $\sigma_L$ $\sim 0.2$ mm. Therefore, $\xi_{Te}$ is dominated by the detector size $L$, i.e.,
 \begin{equation}
     \xi_{Te} \sim \dfrac{\lambda L }{2 \pi \sigma_L} \sim 68.5\, \mathrm{ nm}.
 \end{equation}
Thickness variations in the samples on the length scales larger than the transverse length  will result in incoherent effects, as shown in \fig{fig: coherence setup}.


\subsection{Correction due to bunch spacing incoherence}\label{app: BS incoherence}
\begin{figure}[H]
\centering
  \includegraphics{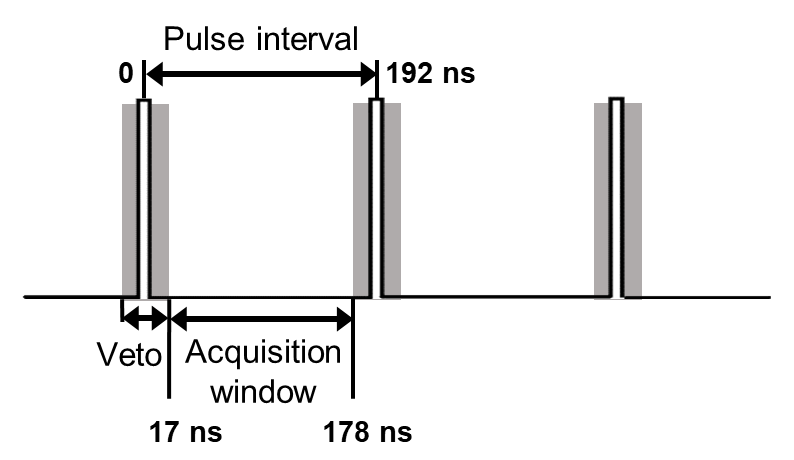}
 \caption{\textbf{Timing structure of the  synchrotron radiation in PETRA-III, 40-bunch mode:} The synchrotron pulses are have a width of 100 ps and are separated by 192 ns. The detectors are given a veto signal to ensure detection occurs only between 17 ns and 178 ns after the X-ray pulse is scattered by the object.}
\label{fig: incoherent_contributions_1}
\end{figure}

In the experimental setup, scattered photons from various X-ray pulses hitting the sample are histogrammed into bins to generate the temporal response. Since the detectors are synchronised with the synchrotron's bunch clock, they always reset to time zero with each repetition of the bunches.
Photons from an earlier X-ray pulse may scatter at times later than $T_\mathrm{b}$, which is the bunch clock's time period. The histogram at time $t_i -T_\mathrm{b}$ contains an incoherent addition of these photons scattered at time $t_i > T_\mathrm{b}$. The histogram with the bunch spacing correction \( I_b \) is thus given by
\[
I_b(t_i) = 
\begin{cases} 
\sum_{k=0}^{N_{b}} I(t_i - k \cdot T_b), & \text{for } t_i \leq T_{b}, \\
0 & \text{otherwise}.
\end{cases}
\]
The bunch clock time period gives the X-ray pulse interval. In our experiment, $T_b \approx 192$ ns, as shown in \fig{fig: incoherent_contributions_1}.

\section{Optimization by Maximum likelihood estimation } \label{sec: max-likelihood}

This section explains the choice of cost function (main text, Eq. 6) for the NuPty phase retrieval algorithm. In the maximum likelihood estimation scheme, we aim to formulate a cost function that incorporates some knowledge of the noise model in the experiment to yield a superior reconstruction. 
In further discussion, we would use the short-hand notation $\boldsymbol{x}_i$ to denote $\boldsymbol{x}(t_i)$, i.e., the value of the vector $x \in \mathbb{C}^{N}$ at the $i$-th time point. The phase problem then becomes as follows: Maximize the joint likelihood function $\mathcal{L}: \mathbb{C}^{N} \rightarrow [0, \infty)$ given as
\begin{equation}\label{eq: likelihood problem}
    \mathcal{L}(\zO) = \prod_{j} \prod_{i} {p}(\zb_{ji}|\zO )
\end{equation}
where  ${p}(\zb_{ji}|\zO)$ is the Bayesian probability of measuring $\zb_{ji}$ counts at the detector at time $t_i = (i-1) \Delta t$ and probe detuning $D_j$, given the object's energy domain response is $\zO$.

The likelihood function is a statistical metric for the consistency of the forward model with the measurements. If the model is changed to make the measurements more probable, the likelihood increases, indicating that the model is better. We try to find an object $\zO$ that maximizes the likelihood $\mathcal{L}$ or, equivalently, minimizes the negative log-likelihood since it is more numerically stable. The cost function is thus given as 
\begin{equation} \label{eq: negloglikelihood}
    \rho (\zO) = -\mathrm{log}(\mathcal{L(\zO)}).
\end{equation} 
The probabilities in \eq{eq: likelihood problem} have to include modeling of all sources of noise in the forward process. Let us assume that each measurement can be modeled as
\begin{equation} \label{eq: noise model}
    \zb_{j} = \mathrm{Poisson}(\boldsymbol{I}_j) + \zn_j
\end{equation}
where $\boldsymbol{I}_j = \left|\mathbf{F} \left\{ \zP_{j}\circo \zO  \right \} \right|^2$, i.e., the noiseless modeled intensity,  $\mathrm{Poisson}(\cdot)$ denotes sampling the measurement from a Poisson distribution based on the modeled intensities and $\zn_j$ is a random noise vector sampled from a Gaussian distribution $\mathcal{N}(0, \boldsymbol{\sigma}_j^2)$. If we suppose that there are no sources of random noise present in the experiment (i.e., $\zn_j =0$), the probabilities can be written with a Poisson distribution model such that
\begin{align}\label{eq: poisson model}
    p(\zb_{ji}|\zO ) = \dfrac{\zI_{ji}^{\zb_{ji}} \cdot \mathrm{e}^{-\zI_{ji}} }{\zb_{ji}!}.
\end{align}
From \eq{eq: negloglikelihood} the cost function to be minimized is given as
\begin{align}
    \rho (\zO) &= -\mathrm{log}\left( \prod_{j}  \prod_{i} p(\zb_{ji}|\zO )\right) \nonumber \\
    &= \sum_{j}  \sum_{i}  -\zb_{ji} \cdot \mathrm{log}( \zI_{ji}) + \zI_{ji} + \mathrm{log}(\zb_{ji}!) . \label{eq: Poisson likelihood}
\end{align}
If we make a Taylor expansion of $\mathrm{log}\left(\sqrt{\zI_{ji}}\right)$ around $\sqrt{\zb_{ji}}$ to the second order, i.e.,
\begin{align}
     \mathrm{log}\left(\sqrt{\zI_{ji}}\right) \approx  \mathrm{log}\left( \sqrt{\zb_{ji}}\right)  +\dfrac{ \sqrt{\zI_{ji}} -  \sqrt{\zb_{ji}} }{ \sqrt{\zb_{ji}}} 
     -\dfrac{\left(\sqrt{\zI_{ji}} -  \sqrt{\zb_{ji}} \right)^2}{ 2{\zb_{ji}}}
\end{align} 
and neglect the constant terms that only depend on $\zb_{ji}$,
we get an approximation of the cost function in \eq{eq: Poisson likelihood} as
\begin{align}
    \rho (\zO) &\approx 2 \sum_{j} \sum_{i}\left( \sqrt{\zI_{ji}} - \sqrt{\zb_{ji} } \right)^2
    = 2 \sum_j  \left\Vert \left|\mathbf{F} 
    \left\{ \zP_j  \circo \zO \right\}\right| - \sqrt{\zb_j}  \right\Vert^2.
    \label{eq: approx Poisson likelihood}
\end{align}  
The Poisson log-likelihood cost function is thus approximately proportional to the {`amplitude-based'} cost function popular in the ptychography community \cite{Yeh2015, Zuo2016}. One advantage of the gradient-based methods is that the phase problem can now be flexibly solved by using the gradient of the cost function in  \eq{eq: Poisson likelihood}. We can calculate the gradient of $\rho$ with respect to the complex object $\zO$ using Wirtinger calculus \cite{Hunger2007} as
\begin{align}
    \nabla_{\zO} \rho &= 2 \sum_j \zP^{\dagger}_j \circo \mathbf{F}^{-1} \left\{\mathbf{F} \left\{ \zZ_j\right\}\right\} \circo \left(  1- \dfrac{\sqrt{\zb_j}}{\left|\mathbf{F} \left\{ \hat{\boldsymbol{Z}}_j \right\}\right|} \right)\nonumber \\
    &= 2 \sum_j \zP^{\dagger}_j \circo \left(\zZ_j -  \mathbf{F}^{-1}\left\{ \dfrac{\sqrt{\zb_j}}{\left|\mathbf{F} \left\{ \hat{\boldsymbol{Z}}_j \right\}\right|} \circo {\mathbf{F} \left\{ \hat{\boldsymbol{Z}}_j \right\}} \right)\right\} \label{eq: grad amp}
\end{align} 
where $\zZ_j = \zP_j \circo \zO$. 
It was shown in Ref.~\cite{Oleh2022} that Rodenburg's PIE algorithm \cite{Rodenburg2004} and its variants are equivalent to the stochastic gradient descent algorithm with the amplitude cost function in \eq{eq: approx Poisson likelihood}. 


\section{Tests on simulated data}

\subsection{Varying Poisson noise}
\begin{figure}[H] 
\centering
\includegraphics{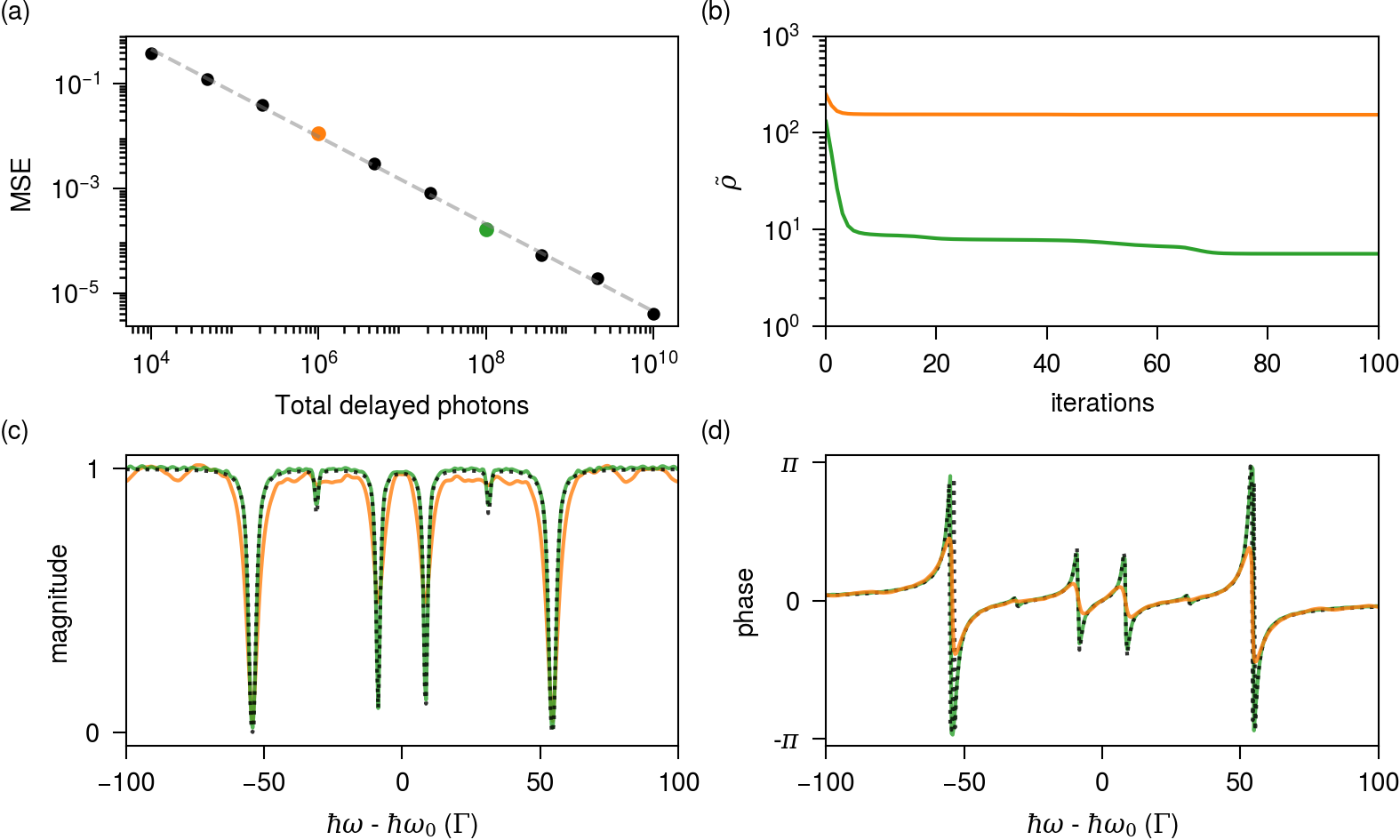}
 \caption{\textbf{Effect of Poisson noise on phase retrieval:}
    (a) Log-log plot of mean squared error (MSE) versus time integrated delayed photons, showing a linear relationship. The trend highlights improved accuracy (lower MSE) as photon counts increase. (b) Convergence of parameter \( \rho \) over 100 iterations, stabilizing after initial decay, with colors indicating photon counts as in (a). Orange and green lines denote results for total delayed photons of \( 10^6 \) and \( 10^8 \), respectively. Higher photon counts help the algorithm reach a deeper cost minimum. (c) Reconstructed magnitude spectrum as a function of energy detuning, \( \omega - \omega_0 \). (d) Reconstructed phase spectrum over the same energy range, showing distinct phase shifts at each resonance line.}
    \label{fig: results_poisson}
\end{figure}

We simulate ptychograms with different levels of Poisson noise by changing the total number of delayed photon counts $N_{ph}$. As described in Sec. \ref{sec: max-likelihood}, the phase retrieval algorithm maximizes the log likelihood of the simulation to the observed data under the assumed Poisson conditions. We always start the first iteration with a priori assumption that our object's nuclei are mostly transparent to X-rays, as the foil's nuclear resonant lines are very sharp. Therefore, we initialize the object $\zO^{(0)}$ as cells with entry 'one'. The object is updated for 100 algorithm iterations with stochastic gradients calculated at $B = 20$. We can determine if the algorithm has reached a solution by tracking the cost $\tilde{\rho}$ per iteration. The performance of the algorithm can be quantified by the mean square error
\begin{equation}\label{eq: MSE}
   \mathrm{MSE} =  \dfrac{1}{N} { \| |\zO^{r}| - |\zO^{*}|\| ^2}
\end{equation} between the magnitudes of the transmission spectra of the final reconstructed object $\zO^{r}$ and the simulated (true) object $\zO^{*}$. In \fig{fig: results_poisson}(a), we see that the MSE decreases as the number of delayed photons in the ptychogram increases and correspondingly the Poisson noise decreases. The algorithm is stable with respect to the increasing levels of Poisson noise. Orange and green colors mark results for total delayed photons of \( 10^6 \) and \( 10^8 \), respectively. In \fig{fig: results_poisson}(b), we show convergence curves of objects reconstructed from ptychograms with the two different levels of Poisson noise. In \fig{fig: results_poisson}(c), we can see that for low photon counts (orange), the reconstructed resonance peaks are broader and the ratio of the heights of the peaks (which corresponds to the relative probabilities of the nuclear transitions) is not preserved. The algorithm reconstructs a "weakly" scattering object, with weaker phase shifts in \fig{fig: results_poisson}(d), for the lower signal to noise ratio. In the ptychogram with higher photon counts, the Poisson noise is low enough that the reconstruction (green) matches perfectly with the true object. In both the low and high noise cases, the major resonance peaks can be identified at their correct locations.

\subsection{Different time windows}
\begin{figure}[htbp!] 
\centering
\includegraphics{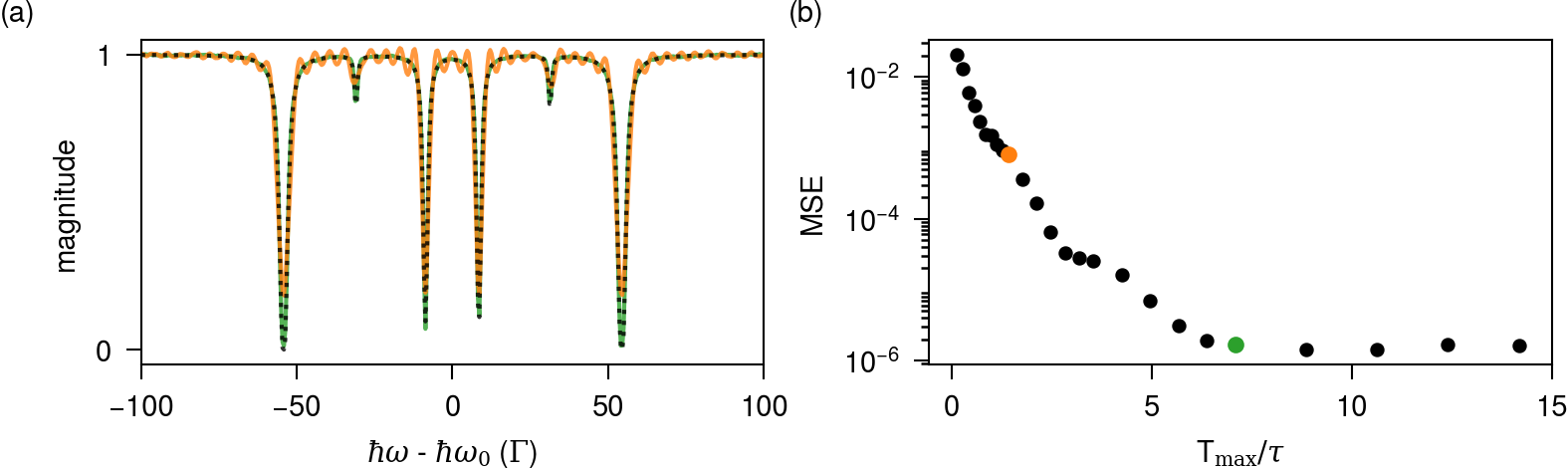}
 \caption{\textbf{Effect of time windowing on phase retrieval:}
(a) Magnitude of the transmission spectrum of the reconstructed object is plotted as a function of energy detuning, \( \hbar \omega - \hbar \omega_0 \). Orange and green lines denote results for $T_\mathrm{max} =$  \( 200 \) ns and \( 1000 \) ns, respectively. For lower $T_\mathrm{max}$, artificial peaks appear in the reconstructed spectrum. (b) Log of mean squared error (MSE) decreases as the maximum acquisition time at the detector (in the unit of the $\Fe$ lifetime $\tau$) increases.}
    \label{fig: results_time_window}
\end{figure}

We now test the effect of the time window cut off $T_\mathrm{max}$ on the phase retrieval. As described in the main text, short time windows of detection cause artificial sinc peaks in the transmission spectrum of the reconstructed object, as seen in \fig{fig: results_time_window}(a). To completely get rid of the artifacts, one needs to increase the time gap between the synchrotron pulses, as seen in the MSE calculations plotted in \fig{fig: results_time_window}(b).

\subsection{Radiative coupling regime in the ptychogram}
\begin{figure}[H]
\centering
\includegraphics[width=0.98\linewidth]{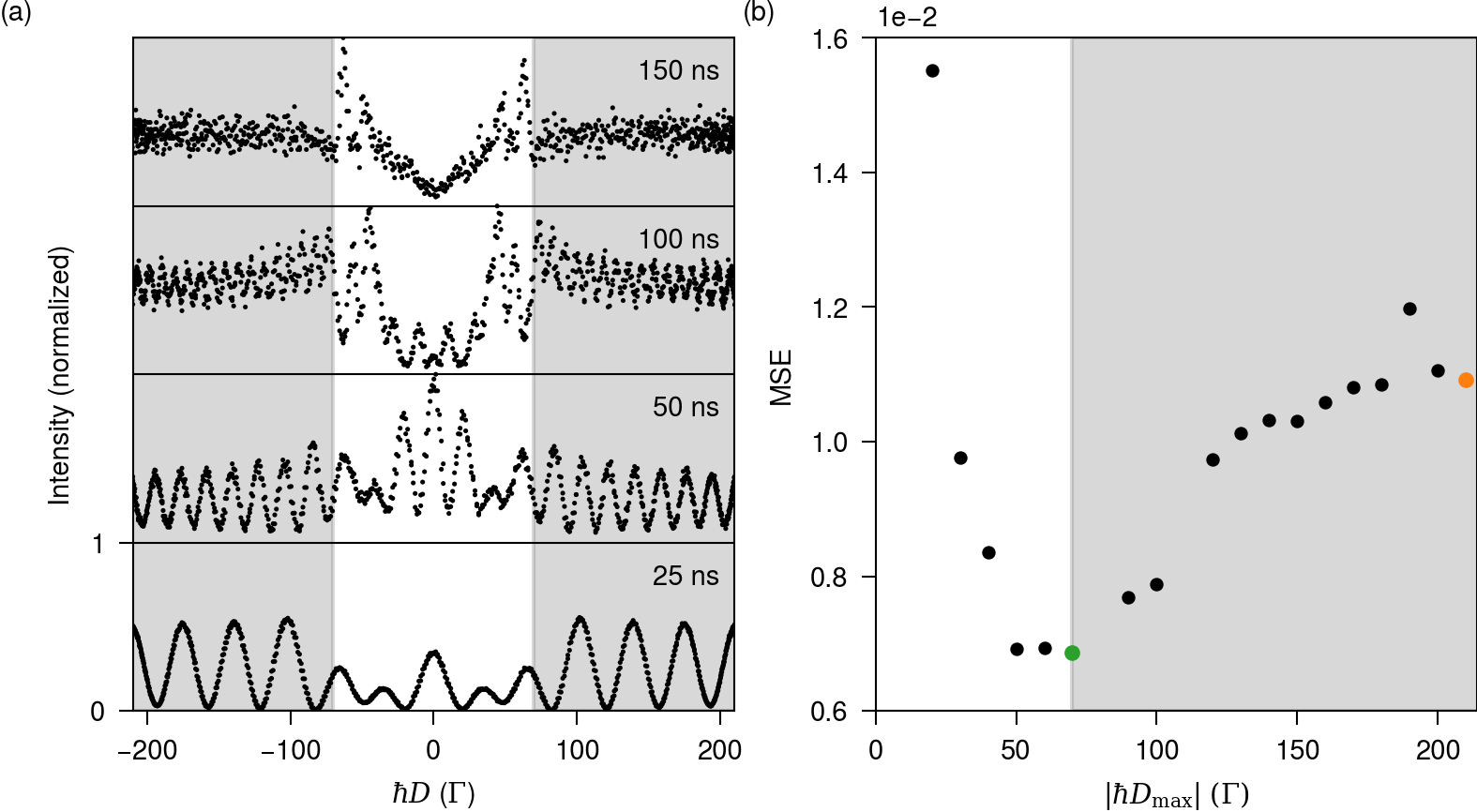}
\caption{\textbf{Effect of probe-object coupling on phase retrieval:} (a) Photon intensity measured at the detector is normalized from 0 to 1 and plotted with respect to Doppler energy detuning $\hbar D$ at different times. (b) The mean squared error (MSE) of the reconstructed object for different maximum Doppler detunings of the probe relative to the object. The orange and green colors correspond to the plots in the main text, Fig. 3(b)-(d). The gray area in both sub-figures marks the interferometric regime ($|\hbar D_\mathrm{max}| > 70$ $\Gamma$) where the coupling between the probe and the object is negligible. }  \label{fig: data_range}
\end{figure}

Neglecting electronic absorption, the total scattering response of the probe and the object can be separated in the time domain as $P(t) = \delta (t) - R_P(t)$ and $O(t) = \delta (t) - R_O(t)$, where $R_P(t)$ and $R_O(t)$ are their respective time delayed nuclear scattered responses.  The response of the probe, Doppler shifted by angular frequency $D = 2 \pi \vvv /\lambda $, is given as
\begin{equation}
    P(D, t) =  \mathrm{e}^{ \imag\varphi(D, t)}{P}(t) = \mathrm{e}^{ \imag Dt}{P}(t)
\end{equation}
and the scattered field at the detector is given by the convolution of their respective temporal responses, i.e.,
 \begin{align} \label{eq: scattered_Z}
 Z(D, t) &=  P(D, t) \ast O(t)\\&=\int_{0}^{t} dt'\,  P(t')  \mathrm{e}^{ \imag D t'}  O(t-t')\\
 &= \delta(t) + e^{\imag D t} R_P(t) + R_O(t) +R_{PO}(D, t)
 \end{align}
The running time $t'$ in the integral is the time of emission of radiation from $P$ and the scattering in $O$. The term $R_{PO} = \int_{0}^{t} dt'\,R_P(t')e^{\imag D t'}R_O(t-t')$ describes the radiative coupling between the probe and the object. It refers to the events where photons get nuclear scattered from both of them such they are coupled by the forward scattered radiation and act as one coherent ensemble.
For very large $D$ the integral $R_{PO} \approx 0$, leading to decoupling of the probe and the object and dominance of interference fringes such that $Z(D, t) \approx \delta(t) + e^{\imag D t} R_P(t) + R_O(t)$. \fig{fig: data_range}(a) illustrates how the measured ptychogram intensities vary with the Doppler detuning, with a clear separation at $\hbar D = \pm 70$ $\Gamma$ between the coupling and interferometric regimes. The ptychography engine produces slightly different reconstructed objects depending on the selected data range from the ptychogram. As shown in \fig{fig: data_range}(b), the mean square error of the reconstruction is minimized when using data with maximum Doppler energy detuning of $\pm 70$ $\Gamma$. 

\section{Regularization techniques for experimental phase retrieval}

\begin{figure}[htbp]
\centering
\includegraphics{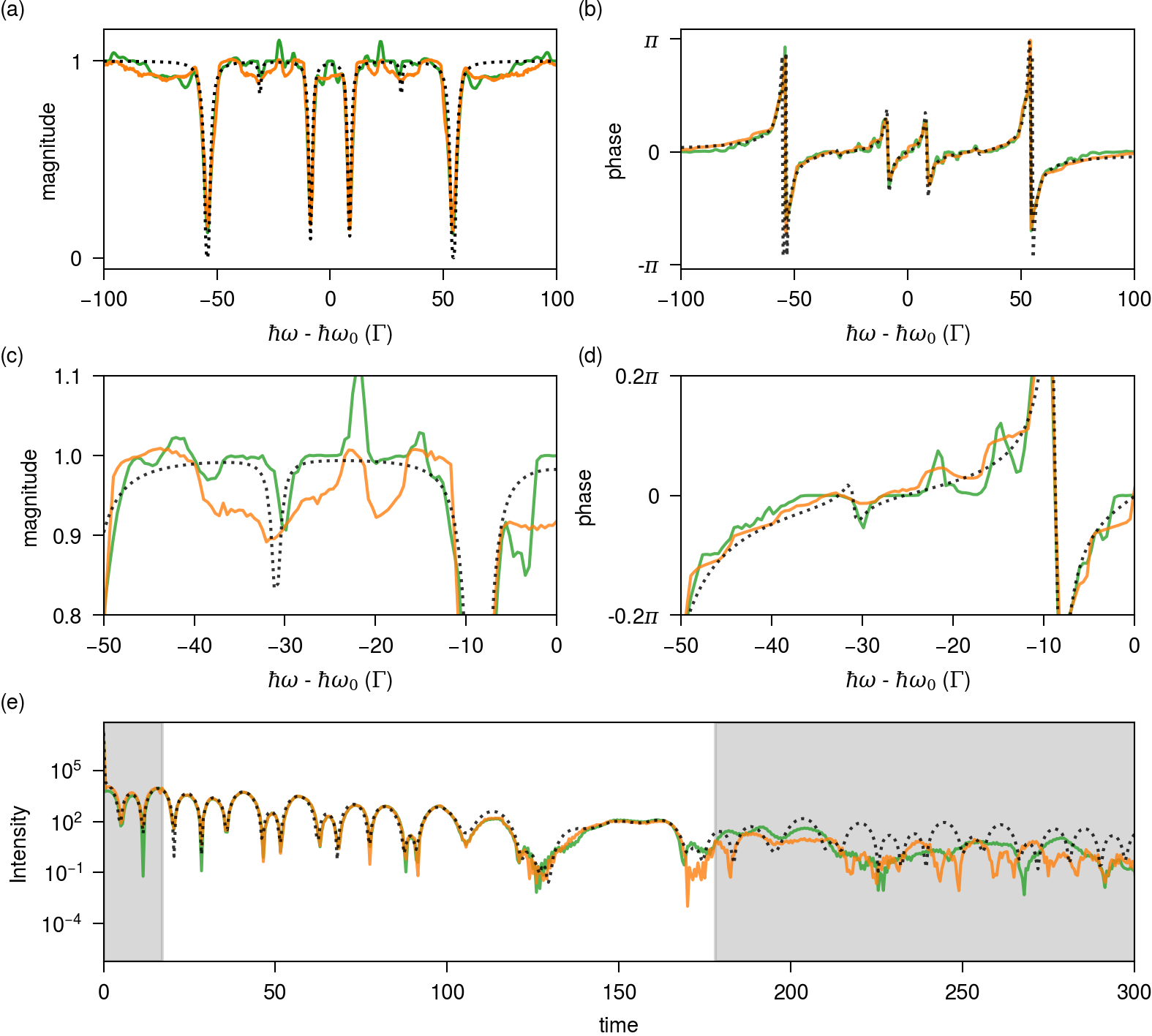}
\caption{\textbf{Effect of regularization on phase retrieval:} Magnitude (a),(c) and phase (b),(d) and (e) time response of the objects reconstructed from experimental data. The simulated object is plotted as black dotted line alongside the total variation regularized reconstruction (orange) and the $\ell^{1}$ regularized reconstruction (green). The gray shaded region in (e) marks the region outside the data acquisition time window (from 17 ns to 178 ns). }  \label{fig: regularization}
\end{figure}

 A standard technique to cosmetically smoothen artifacts in the solution of an inverse problem is to add regularization terms to the cost function, i.e., optimize over $\rho_r(\zO) = \rho(\zO) + r(\zO)$, where $r$ is a carefully chosen regularization or penalty function on the object. In \fig{fig: regularization}(a) and \ref{fig: regularization}(b), we see the magnitude and phase of the object reconstructed from the experimentally measured ptychogram with total variation  regularization $r(\zO) =  \left\Vert\nabla \zO\right\Vert_1 \sim  \left\Vert(\zO)_{i+1} - (\zO)_{i}\right\Vert_1$ (orange) and ( $\ell^{1}$ regularization $ r(\zO) =   \left\Vert 1- \zO\right\Vert_1$ (green). In both cases, the ptychography engine tries to continuously extrapolate the time response beyond $T_\mathrm{max} (= 178$ ns), as seen in \fig{fig: regularization}(e). To some extent, this suppresses the sinc artifacts the reconstructed transmission spectrum (\fig{fig: regularization}(a), (b)). The reconstruction is still not perfect, as one can observe in the zoomed-in plots (\fig{fig: regularization}(c), (d)). This is to be expected since the extrapolated time response beyond $T_\mathrm{max}$ is also artificial. Our regularization constraints relying on the smoothness of the peaks of the energy domain transmission spectrum cannot fully compensate for the missing data in the time domain.


\bibliographystyle{unsrt}  
\newpage
\bibliography{bibliography-file}  

\end{document}